\documentclass[useAMS,usenatbib,fleqn]{mn2e}
\usepackage[dvips]{lscape,graphicx}
\usepackage{url}
\usepackage{amssymb,amsmath}
\usepackage{subfigure}

\begin{document}

\title[A Multi-Wavelength Infrared Study of NGC 891]{A
Multi-Wavelength Infrared Study of NGC 891}
\author[C. H. Whaley, J. A. Irwin, S. C. Madden, F. Galliano and G. J. Bendo]{C. H. Whaley$^{1}$, J. A. Irwin$^{1}$,
S. C. Madden$^{2}$, F. Galliano$^{2, 3}$ and G. J.  Bendo$^{4}$ \\
$^{1}$Queen's University, Kingston, Ontario, K7L 3N6, Canada\\
$^{2}$Service d'Astrophysiques, CEA/Saclay, L'Orme des Merisiers, 91191, Gif-sur-Yvette, France\\
$^{3}$Department of Astronomy, University of Maryland, College Park, MD 20742, USA\\
$^{4}$Astrophysics Group, Imperial College, Blackett Laboratory, Prince Consort Road, London,
SW7 2AZ, United Kingdom}

\date{Accepted year month day. Received year month day; in original form year month day}

\pagerange{\pageref{firstpage}--\pageref{lastpage}} \pubyear{2008}

\maketitle

\label{firstpage}


\begin{abstract}
We present a multi-wavlength infrared study of the nearby, edge-on,
spiral galaxy NGC 891.  We have examined 20 independent, spatially
resolved IR images of this galaxy, 14 of which are newly reduced
and/or previously unpublished images. These images span a wavelength
regime from $\lambda\,1.2$ $\mu$m in which the emission is dominated
by cool stars, through the MIR, in which emission is dominated by
PAHs, to $\lambda\,$850 $\mu$m, in which emission is dominated by
cold dust in thermal equilibrium with the radiation field. The changing morphology of the galaxy with
wavelength illustrates the changing dominant components. We detect
extra-planar dust emission in this galaxy, consistent with
previously published results, but now show that PAH emission is also
in the halo, to a vertical distance of $z\,\ge\,$2.5 kpc.  
We compare the
vertical extents of various components and find that the PAHs (from
$\lambda\,$7.7 and 8 $\mu$m data) and warm dust ($\lambda\,$24
$\mu$m) extend to smaller $z$ heights than the cool dust
($\lambda\,$450 $\mu$m). For six locations in the galaxy for which
the S/N was sufficient, we present SEDs of the IR emission,
including two in the halo --- the first time a halo SED in an
external galaxy has been presented.  We have modeled these SEDs and
find that the PAH fraction, $f_{PAH}$, is similar to Galactic values
(within a factor of two), with the lowest value at the galaxy's
center, consistent with independent results of other galaxies.  In
the halo environment, the fraction of dust exposed to a colder
radiation field, $f_{cold}$, is of order unity, consistent with an
environment in which there is no star formation.  The source of
excitation is likely from photons escaping from the disk.

\end{abstract}

\section{Introduction}
\label{sec:intro}

In a wide variety of galactic environments, polycyclic aromatic
hydrocarbons (PAHs)\footnote{In this paper, we adopt the term "PAH"
to describe the carriers of MIR spectral feature emission, whereas
we refer to thermal continuum emission as dust emission.  Note that
some authors consider PAHs to be low mass carbon grains in their
treatment of interstellar dust (e.g. Draine $\&$ Li 2007b).}
 are the dominant emitters in
the Mid-Infrared (MIR) from about $\lambda\,$5 - 15 $\mu$m$\,$
 (Vogler et al. 2005, Smith et al. 2007, Galliano et al. 2008b). At
 longer wavelengths approaching the Far-Infrared (FIR),
thermal emission dominates the spectral energy distribution (SED)
(Galliano et al. 2008a). PAHs have been observed in many galaxies in
locations often associated with star forming regions. However, new
discoveries of PAH emission extending to significant distances (6 -
10 kpc) from the disks of several galaxies (Irwin \& Madden 2006;
Engelbracht et al. 2006; Irwin et al. 2007) raise important
questions as to how these large
molecules can be present at such high altitudes above the plane and also
their source of excitation. Presumably, far-UV (FUV) emission from in-disk
star forming regions provides the excitation.  However, a fit to a SED
is really required to understand both
the contributions 
from various emission components
(cool stars, PAHs, classical grains, etc.) as well as
the stellar component that is required for the excitation.  Better yet,
a SED fit at different locations in a galaxy, especially at different
locations in the disk and/or in extraplanar gas, can further help us to
 understand
the complex interplay between various IR-emitting
interstellar medium (ISM)
components and the origin
of the excitation.

\begin{figure}
  \includegraphics[scale=0.38]{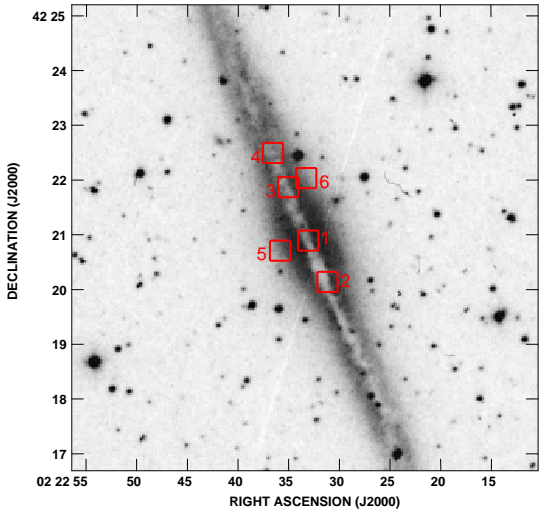}
  \caption{NGC 891 from the Digitized Sky Survey. The numbered red boxes indicate the 
	locations in which the flux was measured to create the spectra.}
  \label{fig:n891_wboxes}
\end{figure}

We have therefore chosen to study MIR-submm emission in a very well
known edge-on galaxy, NGC 891 (Figure \ref{fig:n891_wboxes}), that
has known extraplanar emission
 in a variety of wavebands
(e.g. H I, H$\alpha$, dust and CO - See Sect. \ref{sec:n891}).
Since there is so much spatially resolved
IR data available for this galaxy, it is an excellent target for
modeling its SED at different positions in both the disk and halo regions.
 For this paper, we have examined and compared no less than 20 independent
data sets\footnote{A $\lambda\,160$ $\mu$m image was also examined,
but found to be of too low spatial resolution to be useful.} ranging
in wavelength from $\lambda\,$1.2 $\mu$m to 850 $\mu$m using both
space-based and ground-based instruments
(Sect.~\ref{sec:obs_datared}).  To the limits of the variety of
spatial resolution, coverage, and sensitivity achieved, we compare
the extraplanar extent of the emission and, most importantly, we
derive the SED at different disk and extraplanar locations.



Because of the proximity of NGC~891, its almost exactly
edge-on orientation, its current active star formation (SF) and its
prominent extraplanar emission (see Sect.~\ref{sec:n891}), this
galaxy has been a prime target for studies of the disk-halo
interaction.  
Although the specific vertical height, $z$,
at which gas might be considered in the halo is debatable,
a conservative estimate is 1 kpc (21.5 arcsec).
Since all of the 20 data sets examined have higher
resolution than 21.5 arcsec (see Sect.~\ref{sec:obs_datared}), it is
possible to distinguish the disk from extraplanar (halo) region
 in the observations presented in this paper.

In the next section we describe NGC~891.
In Sect. \ref{sec:obs_datared}, we discuss the
observations and data reduction. The morphology of
NGC 891 is discussed in Sect. \ref{sec:results}, and the SED results
are given in Sect. \ref{sec:sed}. Sects.
\ref{sec:discussion} and \ref{sec:conclusion} present the discussion and
conclusion, respectively.

\subsection{NGC 891}
\label{sec:n891}

NGC~891 (Figure~\ref{fig:n891_wboxes},
Table~\ref{tab:galaxy_params}) is believed to be similar to the
Milky Way (MW) in B-band luminosity, rotational velocity, and Hubble
type. 
It is 2.5 times brighter in the IR, and may have 
twice as much gas and dust than the MW
(see Scoville et al. 1993), although the difference is not strong,
given uncertainties.  There has been much discussion in the literature
as to whether or not NGC~891 is a `starburst' galaxy,
since its star formation rate (SFR) appears to be intermediate between that of a 
normal spiral and a starburst galaxy; it may be in a phase that
saw more powerful starbursting in the past (see Temple et al. 2005).
NGC~891's current SF activity is clearly indicated by a variety of
measures, including
the detection, in H$\alpha$,
of many bubbles, shells, and supershells in the mid-plane of this
galaxy (Rossa et al. 2004).

\begin{table}
 \centering
 \begin{minipage}{140mm}
  \caption{Basic Galaxy Parameters of NGC 891}
  \label{tab:galaxy_params}
  \begin{tabular}{lll}
\hline
parameter & value & reference\\
\hline
Hubble type & SA(s)b & RC3\footnote{RC3 = Third Reference Catalogue by de Vaucouleurs et al. (1991)} \\
RA(J2000) & 02h 22m 33.4s & NED\footnote{NED = NASA Extragalactic Database \\ (\url{http://nedwww.ipac.caltech.edu/})} \\
DEC(J2000) & +$42^\circ$ $20^\prime$ $57^{\prime\prime}$ & NED \\
redshift  & 528 km/s & NED \\
distance & 9.6 Mpc & Strickland et al. (2004) \\
optical major axis & 13.5 arcmin & RC3 \\
optical minor axis & 2.5 arcmin & RC3 \\
major axis angle & $22^\circ$ & NED \\
inclination angle & $89^\circ$ & Baldwin $\&$ Pooley (1973) \\
rotational velocity & 225 km/s & Sofue (1997) \\
B band luminosity & 7.8 $\times$ $10^{9}$ $L_\odot$ & RC3 \\
SFR & 3.8 $M_\odot$/yr & Popescu et al. (2004) \\    
$M_{gas}$ (H I $\&$ $H_2$) & 4.6 $\times$ $10^9$ $M_{\odot}$ & Guelin et al. (1993) \\
 & & Dupac et al. (2003) \\
$M_{dust}$ & 1.9 - 7 $\times$ $10^7$ $M_{\odot}$\footnote{The range reflects different
results from various authors.} & Alton et al. (2000) \\
 & & Popescu et al. (2004) \\
 & & Galliano et al. (2008a) \\
$M_{PAH}$ & 3.3 $\times$ $10^6$ $M_{\odot}$ & Galliano et al. (2008a) \\
\hline
\end{tabular}
\end{minipage}
\end{table}

The molecular gas, as observed in CO,
 resembles that of the MW and is mainly confined to a
thin disk (400 pc thickness) (Scoville et al. 1993) although a halo
component is also observed (see below). 
Emission from cold dust (average temperature of $\approx$ 18 - 24 K;
Dupac et al. 2003), as observed at $\lambda\,$1.3 mm, 850 $\mu$m and
450 $\mu$m, correlates spatially with the molecular gas in the disk
(Alton et al. 1998; Israel 1999). Spatial correlations have also
been found between cold dust at $\lambda\,$850 $\mu$m, the 7.7
$\mu$m PAH feature and very small grain (VSG) emission observed at
$\lambda\,$14.3 $\mu$m that is not specifically associated with
starburst regions (Haas et al. 2002).


As indicated in Sect.~\ref{sec:intro}, NGC~891 has extended extraplanar
gas in a variety of ISM components.
The galaxy has widespread diffuse ionized gas (DIG) in its halo
(Rossa $\&$ Dettmar, 2003; Rossa et al. 2004)
up to 2.2 kpc above the galactic plane. It is one of the brightest
of all nearby edge-on spirals, suggesting pronounced
disk-halo mass transfer (Alton et al. 2000).  The H$\alpha$ halo emission
is more prominent and extended in the $z$ direction
on the northern side of the galaxy than the southern.
Some of this asymmetry may be explained by dust obscuration
(Kamphuis et al. 2007), but Rossa et al. (2004) interpret
this as being due to a higher SFR in the northern part of the disk
than in the southern part. The assymetry is also seen in the 
radio continuum observations (Dahlem et al. 1994) and these are 
not affected by dust. They therefore, support the latter interpretation.

The broadest halo component is that of H I, which also shows a
north-south asymmetry in disk thickness (Swaters et al. 1997), and
is now known to extend up to at least 22 kpc above the galactic
plane (Oosterloo et al. 2007). Also, the H I halo is asymmetric: It is 
clearly thicker in the north than in the south of the galaxy (Swaters et al. 1997).
Molecular gas, as observed in CO, has been detected to 1 - 1.4 kpc above the plane
(Garcia-Burillo et al. 1992).  As for dust in the halo, optical absorption
features have been detected to heights of 2 kpc
(Howk $\&$ Savage 2000). 
Burgdorf et al. (2007) have also
detected $\lambda\,$16 and $\lambda\,22$ $\mu$m 
emission extending up to 5 kpc from the midplane.  Recently, Rand
et al. (2008) have obtained IR spectra showing the [NeII] and [NeIII]
MIR lines of two positions in NGC~891's
halo which suggest that the radiation field may harden with height
from mid-plane.

\section{Observations and Data Reduction}
\label{sec:obs_datared}

This paper makes use of both archival data that we have reduced
as well as previously published data.  In this section,
we first describe the data that we reduced ourselves and then the origin of
the data sets that were obtained from others.

\subsection{ISO Observations and Data Reduction}
\label{sec:iso_obs}

The Infrared Satellite Observatory (\textit{ISO}) archive was accessed to
obtain two previously unpublished data sets that were
acquired in 1997 and 1998. 
 The data were taken with \textit{ISO}'s camera,
ISOCAM, using the long wave channel, with  narrow band filters,
LW1, LW4, LW5, LW6, LW7, LW8, LW9, (centered at
$\lambda\,$4.5, 6.0, 6.8, 7.7, 9.6, 11.3, 14.9
$\mu$m, respectively), and wide band filters, LW2 and LW3 ($\lambda\,$6.7
and 14.3 $\mu$m, respectively). 
All of these observations were with the CAM01 observing mode, which is
the standard mode for photometric imaging (Blommaert et al. 2003).

The galaxy was not completely contained in the field of view. Only
3.2 arcmin $\times$ 3.2 arcmin (8.9 $\times$ 8.9 kpc) were viewed at
a time, and the galaxy was not always centered in the field. There
were multiple pointings which had to be mosaiced together (see
below) to form the maps shown in Sect. \ref{sec:iso_images}. The raw
data consist of a sequence of 2-dimensional frames taken at
successive times for any given wavelength band, forming a `cube' of
data.  Table \ref{tab:obs_params} summarizes the observation
parameters.

\begin{landscape}
 \begin{table}
 \begin{minipage}{300mm}
  \caption{ISO observing and map parameters.}
  \label{tab:obs_params}
{\small
  \begin{tabular}{llllllllll}
  \hline
 & \multicolumn{9}{c}{wave bands} \\
Parameter & LW1 & LW4 & LW2(wide) & LW5 & LW6 & LW7 & LW8 & LW3(wide) & LW9 \\
\hline
central wavelength\footnote{Blommaert et al. (2003)} ($\mu$m) & 4.5 & 6.0 & 6.7 & 6.8 & 7.7 & 9.6 & 11.3 & 14.3 & 14.9 \\
wavelength range$^a$ ($\mu$m) & 4.0-5.0 & 5.5-6.5 & 5.0-8.5 & 6.5-7.0 & 7.0-8.5 & 8.5-10.7 & 10.7-12.0 & 12.0-18.0 & 14.0-16.0 \\
TDT number\footnote{a unique number that identifies the observation.} & 65101201 & 65101201 & 65101201  & 65101201 & 65101201 & 65101201 & 65101201 & 65101201 & 65101201 \\
 & & & 84101766 & & & & & 84101766 & \\
pixel field of view\footnote{See text Sect. \ref{sec:iso_obs}} (arcsec) & 6.0 & 6.0 & 6.0 & 6.0 & 6.0 & 6.0 & 6.0 & 6.0 & 6.0 \\
PSF\footnote{The point spread function (PSF) full width at half maximum (FWHM).} (FWHM) (arcsec) & 7.3 & 7.3 & 7.3 & 7.9 & 7.9 & 7.9 & 7.9 & 7.9 & 8.4 \\
date of observation & 28 Aug 97 & 28 Aug 97 & 28 Aug 97 & 28 Aug 97 & 28 Aug 97 & 28 Aug 97 & 28 Aug 97 & 28 Aug 97 & 28 Aug 97 \\
 & & & 5 March 98 & & & & & 5 March 98 & \\
no. of on-source pointings\footnote{This is the number of different pointing positions on-source.} & 7 & 7 & 15 & 7 & 7 & 7 & 7 & 15 & 7 \\
mean $\#$ frames/pointing & 16 & 16 & 13 & 16 & 13 & 16 & 16 & 16 & 16 \\
integration time per frame (s) & 10.08 & 10.08 & 10.08 & 10.08 & 10.08 & 10.08 & 10.08 & 10.08 & 10.08 \\
total on source time\footnote{Total integration time per on-source pointing times the number of on-source pointings.} (min) & 18.8 & 18.8 & 36.8 & 18.8 & 15.3 & 18.8 & 18.8 & 40.3 & 18.8 \\
sky coverage (arcmin) & 3.2 & 3.2 & 3.2 & 3.2 & 3.2 & 3.2 & 3.2 & 3.2 & 3.2 \\
RMS of final maps (MJy/sr) & 0.12 & 0.14 & 0.13 & 0.22 & 0.16 & 0.10 & 0.28 & 0.19 & 0.19 \\
dynamic range of final maps\footnote{maximum of map divided by the RMS (noise) of final map.} & 189 & 202 & 370 & 167 & 600 & 119 & 138 & 136 & 91 \\
\hline
 \end{tabular}
}
\end{minipage}
 \end{table}
 \end{landscape}

For the data reduction, we used the
Cam Interactive Reduction (CIR) package (Chanial 2003)
which runs in Interactive Data Language (IDL).  A thorough review
of the data reduction procedures can be found in Galliano (2004).

We first performed a dark correction following the characterization
of Biviano et al. (1998).  A second-order correction is also
carried out during this step, depending on the detector temperature
and the time since activation, and a
 short-drift correction is applied. 

Cosmic rays were then removed by running an automatic
multi-resolution de-glitching routine which
masks the glitches that are short in duration and have an intensity above
6$\sigma$ (where $\sigma$ is the noise level).
Further manual de-glitching was then carried out, frame by frame,
 to remove the remaining
glitches that did not meet these criteria.  These were typically
pixels that were immediately
adjacent to ones that had been automatically deglitched.

Memory effects caused by non-instantaneous temperature changes in
the photoconductors were then corrected using the
Fouks-Schubert (1995) method and the Wozniak implementation
(see Galliano 2004).

We removed a narrow border of each frame, since the pixels at the
very outer edges of the CCD have higher noise. This step has not 
significantly reduced the field of view because the various
individual frames overlap spatially and the final field of view
is dictated by the mosaic of these frames.

The data were separated into blocks of frames corresponding to each
wave band. The blocks were then registered to the same sky position
and medianed together to create a single image at each wave band.

We then flat fielded each image using flat fields maintained in a
flat library at the ISO Data Center and also corrected for flat
distortion which is a known ISOCAM error. At this point, the data
were calibrated into mJy/pixel units using the calibration terms
given by Blommaert et al. (2003).

Background emission was measured in a region that was far from the
main emission (greater than 15 kpc).  This background emission was
then subtracted from the image.


Finally, we created \textit{error maps} for each wave band (not
shown), which represent pixel-by-pixel RMS values derived during
cube contraction, propagating through errors in the various
processing steps.

The remaining source of error not included in the error maps is the
ISOCAM absolute flux calibration error, which does not affect the
pixel-by-pixel variation within a given image. The calibration error
is by far the largest of the errors, at about 20$\%$ (Pagani et al.
2003; Coia et al. 2005). 

The effect of a dead pixel column on the detector remains on all of
our ISOCAM images. This region is blanked in Figures
\ref{fig:pah_cont}, \ref{fig:Apah}, \ref{fig:Aother}
and \ref{fig:Awide}. We do not make any measurements in regions 
affected by this known error.

\subsection{Spitzer Observations and Data Reduction}
\label{sec:MIPS_obs}

We use \textit{Spitzer} Space Telescope data with both the Multiband
Imaging Photometer (MIPS) as well as the Infrared Array Camera
(IRAC).  Observing and map parameters for the Spitzer data can be
found in Table~\ref{tab:obs_spitzer}.


MIPS scan map data were only initially available for Astronomical
Observation Requests 14815744, 14815488, and 14816000, which were
data taken as part of an observing program led by C. Martin, so only
those data were used in our analysis.  The raw data were processed
and combined into final mosaics using the MIPS Data Analysis Tools
version 3.06 (Gordon et al. 2005).  The data processing for the
$\lambda\,$24~$\mu$m data differs significantly from 
the $\lambda\,$70~$\mu$m data, so the
data reduction procedures are given separately.

The individual $\lambda\,$24~$\mu$m frames were first processed through a droop
correction (removing an excess signal in each pixel that is
proportional to the signal in the entire array) and were corrected
for non-linearity in the ramps.  The dark current was then
subtracted. Next, scan-mirror-position dependent flats were created
from the data in each 
Astronomical Observation Request (AOR)
and were applied to the data to remove dark
spots in the image that are caused by contamination on the MIPS scan mirror.
Scan-mirror-position independent flats were created from the data in 
each AOR and were applied to the data to remove negative latent images from observations 
of bright objects that preceded the individual AORs and to correct sensitivity 
variations across the detector.
Following this, the
zodiacal light emission was fit with a third-order polynomial
function in each scan leg, and the function was then subtracted from
the data.  A robust statistical analysis was then applied in
which the values of cospatial pixels from different frames were
compared to each other, and statistical outliers such as cosmic rays
were masked out.  After this, a final mosaic was made with pixel
sizes of 1.5~arcsec, any residual background in the image was
subtracted, and the data were calibrated into astronomical units
using the calibration term 0.0447 MJy sr$^{-1}$ $MIPS_{unit}^{-1}$.
The uncertainty in the calibration is 4\%.  While the FWHM of the
PSF is $\sim6$~arcsec, significant extended emission from the Airy
rings can be found outside the central peak.

In the $\lambda\,$70~$\mu$m data processing, the first step was to fit ramps to
the reads to derive slopes.  In this step, readout jumps and cosmic
ray hits were also removed, and an electronic nonlinearity
correction was applied.  Next, the stim flash frames taken by the
instrument were used as responsivity corrections.  The dark current
was subtracted from the data, and an illumination correction was
applied.  Short term variations in the signal (often referred to
as drift) were removed from the data, which also subtracted the
background from the data. Next, a robust statistical analysis was
applied to cospatial pixels from different frames in which
statistical outliers were masked out. Once this was done, final
mosaics were made using square pixels of 4.5~arcsec.  The residual
backgrounds were measured in regions outside the optical disk of the
galaxy and subtracted, and then the flux calibration factor of 0.2
MJy sr$^{-1}$ $MIPS_{unit}^{-1}$ was applied to the data.  The
uncertainty in the calibration term was 7\%. 
The final image is very strongly affected by bright and dark streaks
outside the plane of the galaxy that are related to latent image
effects. Also, while the FWHM of the PSF is $\sim18$~arcsec,
significant extended emission from the Airy rings can be found
outside the central peak.

Aside from uncertainties in the MIPS calibration terms indicated above,
there are additional non-linearities and other effects which we estimate
puts errors on flux measurements, when comparing with other bands, of
order 10 to 15\%. 

The IRAC observations at $\lambda\,$3.6, 4.5, 5.8, and 8.0 $\mu$m
were taken from the \textit{Spitzer} archive. These are the
post-basic calibrated data (data that are already reduced and
calibrated) in four MIR wave bands: $\lambda$\,3.6, 4.5, 5.8, and
8.0 $\mu$m. These data were part of the Brown Dwarf Galaxy Haloes
observing program led by Giovanni Fazio (Program ID \#3).
Calibration uncertainties are of the order of 5\% (Reach et al. 2005).

Additional data reduction on these IRAC data were as follows: First, a
sky subtraction 
was performed on each band.
Then, maximum `infinite aperture' corrections of 0.94, 0.94, 0.78, and
0.74 have been recommended for the $\lambda\,$3.6, 
$\lambda\,$4.5, $\lambda\,$5.8, and $\lambda\,$8.0 wavebands,
respectively (Reach et al. 2005) for extended sources. Since these
corrections are of order of the absolute calibration error for the 3.6
and 4.5 micron bands, we make no corrections for those bands.
For the longer wavelength bands, however, we have applied a correction
for all disk emission flux measurements 
in the amount of 0.94 and 0.78 at $\lambda\,$5.8
and $\lambda\,$8.0, respectively. These values assume that the maximum error
must be applied in a direction along the disk whereas the disk is
well defined and finite in the z direction (see Figure
\ref{fig:77minor_axis} plus the Spitzer website\footnote{\url{
http://ssc.spitzer.caltech.edu/irac/calib/extcal/}}).  
Since infinite
aperture corrections are unknown for the halo regions, we omit the
5.8 and 8.0 $\mu$m data points in the SED halo calculations (Section
\ref{sec:sed_results}).  The displayed maps are shown without this 
correction.

For additional IRAC details and map rms, we refer to Table
\ref{tab:obs_spitzer}.

\subsection{2MASS and JCMT Observations}
\label{sec:other_obs}

The remaining data sets include the 2 Micron All Sky Survey (2MASS)
images in the J, H and K bands, which were obtained from the Skyview
archive\footnote{\url{http://skyview.gsfc.nasa.gov/cgi-bin/skvadvanced.pl}}
 (Jarrett et al. 2003), as well as the Sub-millimeter Common User Bolometer Array
 (SCUBA) data of the James Clerk Maxwell Telescope (JCMT) at
$\lambda\,$450 and 850 $\mu$m which were kindly supplied by E.
Xilouris (see Alton et al. 1998). Calibration uncertainties are
approximately 2$\%$ for 2MASS (Cohen et al. 2003) and 15 - 25$\%$
for SCUBA (Alton et al. 1998 and Tilanus 2005).

For the SCUBA data, we adopt the estimates of Israel et al. (1999) who indicate
that about 5\% CO(3-2) contamination may be in the $\lambda\,$850
$\mu$m band with $<\,$10\% toward the nuclear region, and that negligible
CO(6-5) contamination is in the $\lambda\,$450 $\mu$m band.  Since
these are less than the calibration uncertainties, we do not correct
for CO in these bands.

For observation and map parameters for these data, see Table
\ref{tab:obs_other}.

\subsection{Final Maps}
\label{sec:final_maps}

For comparative purposes, all data were regridded to the same pixel
size (1 arcsec) and the flux converted to units that could be
compared, map to map.  The units are therefore, either mJy
arcsec$^{-2}$ or MJy sr$^{-1}$ (note that the conversion from mJy
arcsec$^{-2}$ to MJy sr$^{-1}$ requires a multiplication by 42.55).
Most data in this paper are presented in MJy sr$^{-1}$ units, though
we have also retained the original Jy beam$^{-1}$ units for the
JCMT/SCUBA data to present those maps in a form that is standard at
those wavelengths.


\begin{landscape}
 \begin{table}
 \begin{minipage}{300mm}
  \caption{Spitzer observing and map parameters.}
  \label{tab:obs_spitzer}
  \begin{tabular}{lllllll}
  \hline
Parameter\footnote{See Table \ref{tab:obs_params} for explanation of
parameters
that are not described here.} & Spitzer/IRAC & & & & Spitzer/MIPS &  \\
\hline
reference wavelength\footnote{The reference wavelength is the name for the wave band that is often used in the literature.} ($\mu$m) & 3.6 & 4.5 & 5.8 & 8.0 & 24 & 70  \\
central wavelength\footnote{The central wavelength is the
actual center of the wave band.} ($\mu$m) & 3.6 & 4.5 & 5.8 & 8.0 & 23.68 & 71.42 \\
bandwidth\footnote{Wavelength range of filter.} ($\mu$m) & 1.0 & 1.0 & 1.5 & 3 & 8 & 20  \\ 
pixel field of view (arcsec) & 1.2 & 1.2 & 1.2 & 1.2 & 1.5 & 4.5  \\
PSF (FWHM) (arcsec) & 1.9 & 2.0 & 1.9 & 2.2 & 6 & 18  \\
date of observation & 8 Sept 2004 & 8 Sept 2004 & 8 Sept 2004 & 8 Sept 2004 & 4 Sept 2005 & 4 Sept 2005  \\
RMS of final maps (MJy/sr) & 0.026 & 0.018 & 0.023 & 0.057 & 0.048 & 0.050  \\
dynamic range of final maps & 136 & 169 & 309 & 611 & 587 & 490  \\
\hline
\end{tabular}
\end{minipage}
\end{table}

 \begin{table}
 \begin{minipage}{300mm}
  \caption{2MASS and JCMT/SCUBA observing and map parameters.}
  \label{tab:obs_other}
  \begin{tabular}{lllllll}
  \hline
Parameter & 2MASS & & & JCMT/SCUBA &  & \\
\hline
waveband\footnote{Name of wave band.} ($\mu$m) & J & H & K & 450 $\mu$m & 850 $\mu$m &  \\
central wavelength ($\mu$m) & 1.24 & 1.66 & 2.16 & 442 & 862 &  \\
bandwidth$^d$ ($\mu$m) & 0.162 & 0.251 & 0.262 & 50 & 130 & \\ 
PSF (FWHM) (arcsec) & 1 & 1 & 1 & 7.5 & 15.7 &   \\
RMS of final maps & 1.7 $\times$ $10^{-7}$ MJy/sr & 3.1 $\times$
$10^{-7}$ MJy/sr & 2.8 $\times$ $10^{-7}$ MJy/sr
& 0.83 Jy/beam & 0.010 Jy/beam & \\
\hline
\end{tabular}
\end{minipage}
\end{table}
\end{landscape}


\section{Results}
\label{sec:results}

\subsection{IR Images and Emission from the Disk}
\label{sec:iso_images}

\begin{figure}
  \includegraphics[scale=0.45]{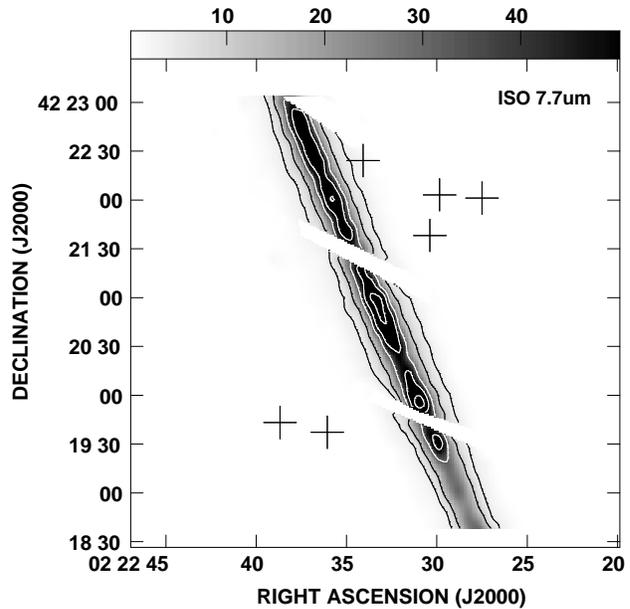}
  \includegraphics[scale=0.45]{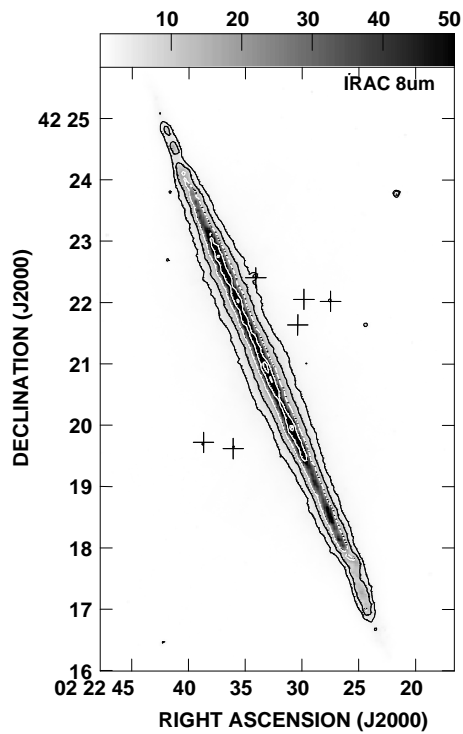}
\caption{Surface brightness contours of the ISO $\lambda\,$7.7 $\mu$m and the IRAC $\lambda\,$8 $\mu$m wave 
bands that select PAH emission. The crosses denote foreground stars. The contours are 0.4 (3$\sigma$), 
1, 3, 5, 8 MJy/sr for $\lambda\,$7.7 $\mu$m, and 0.17 (3$\sigma$), 0.4, 1.3, 5, and 7 MJy/sr for $\lambda\,$8 $\mu$m.} 
\label{fig:pah_cont}
\end{figure}

\begin{figure}
  \includegraphics[scale=0.4]{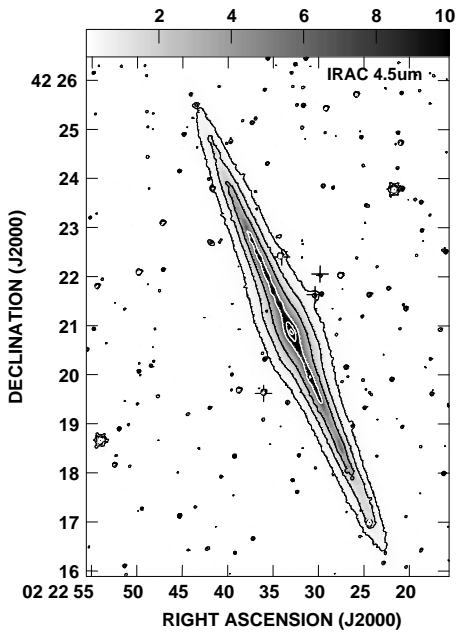}
  \caption{Surface brightness contours of the IRAC $\lambda\,$4.5 $\mu$m wave band that selects mostly the 
	stellar continuum. Symbols as in Figure \ref{fig:pah_cont}. Contours are 0.05 (3$\sigma$), 
	0.15, 0.4, 1.2, 2.5, and 5 MJy/sr.}
    \label{fig:stellar_cont}
\end{figure}

\begin{figure}
  \includegraphics[scale=0.4]{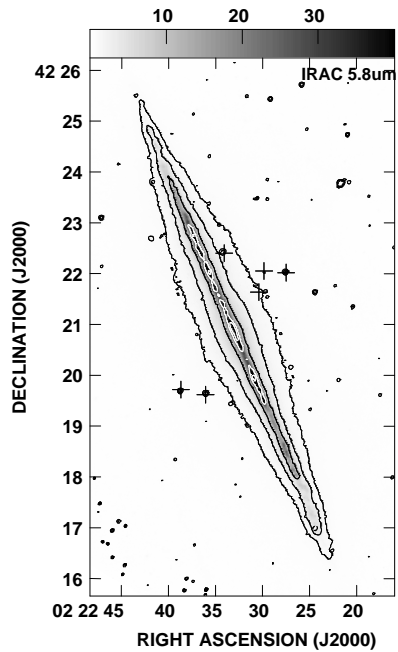}
  \caption{Surface brightness contours of the $\lambda\,$5.8 $\mu$m wave band that selects the MIR continuum. Symbols as in Figure
  \ref{fig:pah_cont}. Contours are 0.068, 0.17, 0.6, 2, 3.4 MJy/sr.}
    \label{fig:other_cont}
\end{figure}

Figures \ref{fig:pah_cont} -
\ref{fig:other_cont} and \ref{fig:Apah} - \ref{fig:Awide} show the MIR surface
brightness of NGC 891 in both contours and grey scale. The ISO $\lambda\,$7.7 $\mu$m
narrow wave band and the IRAC $\lambda\,$8 $\mu$m wave band are particularly good at selecting PAH
spectral features and are shown in Figure \ref{fig:pah_cont} (with the other ISO PAH bands shown in
Figure \ref{fig:Apah}). The $\lambda\,$4.5 $\mu$m
wave band is shown in Figure \ref{fig:stellar_cont}, and the $\lambda\,$5.8 $\mu$m wave band is shown in
Figure \ref{fig:other_cont}. Figure \ref{fig:Awide} shows the two
wide bands from ISO. In each figure the foreground stars have been indicated by small
crosses\footnote{Foreground stars were identified using the 2MASS K
band map.} and the dead pixel column is contained within the blanked 
white region of the ISO images.

\begin{figure}
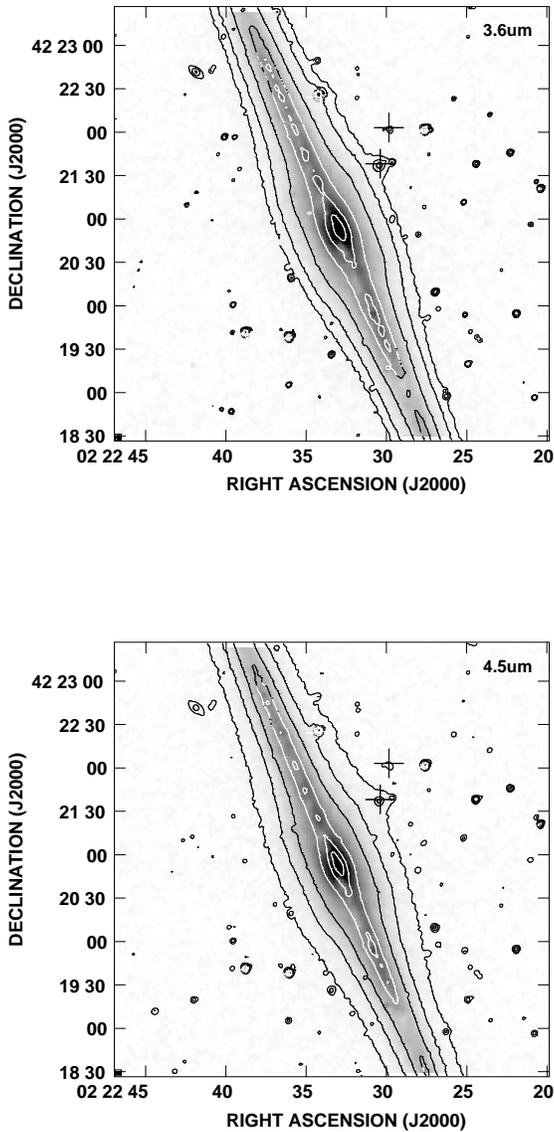

 \includegraphics[scale=0.4]{36_2MASSK.PS}
 \includegraphics[scale=0.4]{IRAC45_2MASSK.PS}
 \caption{IRAC
$\lambda\,$3.6 and 4.5 $\mu$m emission contours overlayed on the K band image. The contours
 are the same as for Figure \ref{fig:stellar_cont}.}
 \label{fig:45on2massk}
\end{figure}

Figures \ref{fig:pah_cont} - \ref{fig:other_cont} (and \ref{fig:Apah} 
- \ref{fig:Awide}) show that most of
the emission is coming from the disk of the galaxy, especially from
the location of the central dust lane. These surface brightness maps
show prominent peaks toward the nucleus and secondary maxima  on
either side of the center. These secondary peaks are associated with 
spiral arms and bright H II regions (Kamphuis et al. 2007). Figure
\ref{fig:stellar_cont} shows that the contours at $\lambda\,$4.5
$\mu$m form a bulge at the center, whereas at longer
wavelengths, we do not see this bulge. Comparison of major axis
profiles (not shown) also reveal substantial differences between the
short and long wavelength emission. This change in morphology with
increasing wavelength illustrates the changing dominance of spectral
components.  The $\lambda\,$3.6 $\mu$m wave band contains the 3.3
$\mu$m PAH feature and Le Coupanec et al. (1999) concluded that the
$\lambda\,$4.5 $\mu$m flux in NGC 891 contained a strong warm dust
component. However, Figure \ref{fig:45on2massk}, which shows the emission contours in these two
wave bands overlaid on a stellar map (2MASS K band), shows an
excellent correspondence between all three bands. This suggests that
the $\lambda\,$3.6$\mu$m and 4.5 $\mu$m bands are dominated by
starlight, a result that is borne out by our SED results
(Sect.~\ref{sec:sed_results}). We are therefore in agreement with
most of the results from SINGS and other Spitzer studies, which have
demonstrated that the 3.6 and 4.5 $\mu$m emission is almost entirely
stellar in origin (Draine et al. 2007b; Smith et al. 2007).

\begin{figure}
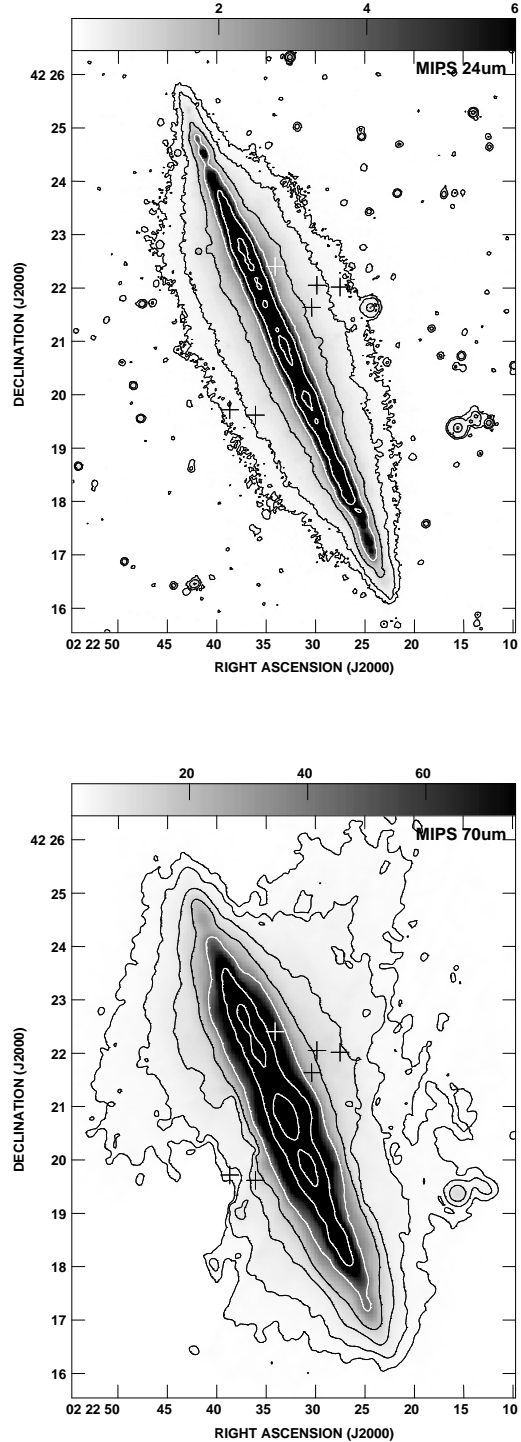

  \includegraphics[scale=0.4]{N891_24MUMIPS.PS}
  \includegraphics[scale=0.4]{N891_70MUMIPS.PS}
  \caption{MIPS
$\lambda\,$24 $\mu$m and 70 $\mu$m maps. The contours start at 3$\sigma$.
    Symbols as in Figure \ref{fig:pah_cont}.
    Contours are 0.145, 0.3, 1, 3, 10, 50 MJy/sr for $\lambda\,$24 $\mu$m, and
    1.5, 3, 7, 15, 40, 120, 350 MJy/sr for $\lambda\,$70 $\mu$m.}
  \label{fig:dust_maps}
\end{figure}

Figures \ref{fig:dust_maps} and \ref{fig:cool_dust_map} show the
warm to cold dust emission as wavelength increases from
$\lambda\,$24, through 70, 450, and 850 $\mu$m. Comparing these maps with
Figures \ref{fig:pah_cont} - \ref{fig:other_cont} and \ref{fig:Apah} - \ref{fig:Awide},
we find that, along the plane of NGC 891, the local
maxima of emission for PAH-sensitive bands are at similar locations as the local
maxima for both warm dust and cold
dust.  
Therefore, to first order, PAHs and dust emission correlate in the
disk. This result is likely simply a statement that where there is a
density enhancement of dust, there will also be a density
enhancement of molecular material and PAHs as well. The dust
temperature and source of PAH excitation, however, impose additional
important constraints on the apparent distribution of these
components.  PAHs and cold dust are known to be correlated from
previous work (Haas et al. 2002; Bendo et al. 2008), suggesting that
PAH excitation is possible from a widespread stellar FUV-emitting
population, rather than in SFRs alone.

\begin{figure}
  \includegraphics[scale=0.4]{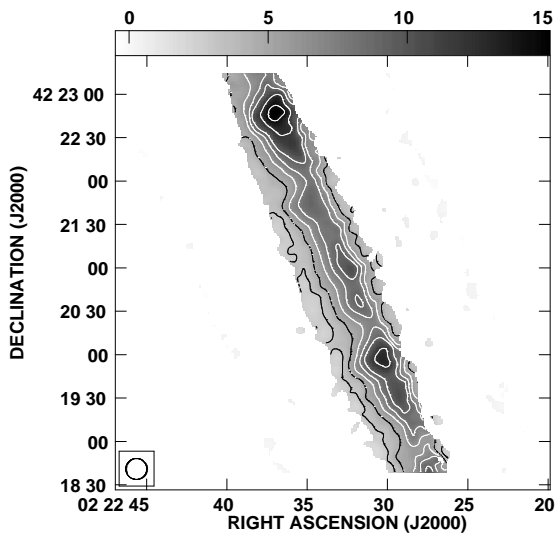}
  \caption{The
$\lambda\,$24 $\mu$m to 850 $\mu$m surface brightness ratio map. The contours are 2, 3, 4, 5, 6, 8, 9.5.}
  \label{fig:24_850_ratio}
\end{figure}

As for warm and cold dust, we investigate the relationship between these
components via the $\lambda\,$24 $\mu$m to 850 $\mu$m surface
brightness ratio map, after first smoothing the
$\lambda\,$24 $\mu$m map to the same resolution as the $\lambda\,$850 $\mu$m map.
The result is shown in Figure \ref{fig:24_850_ratio}.
From this figure, it seems that the cold dust emission ($\lambda\,$850 $\mu$m) is slightly more
extended, since the ratio tends to decrease outward. This agrees
with our results for the $z$ distribution as well (see
Sect.~\ref{sec:minor_axis}). In addition, in the disk there are two
``blobs", or maxima, in the ratio, on either side of the centre
identifying regions in which the dust is hotter than average. The
southern maximum is located at the same position as the surface
brightness maxima seen in the previous maps (Figures
\ref{fig:pah_cont} - \ref{fig:45on2massk} and \ref{fig:cool_dust_map}).
The northern ratio maximum is located further away from the
centre, at a declination of 42$^{\circ}$ 22$^{\prime}$
45$^{\prime\prime}$. Both these maxima are located in regions in
which the H$\alpha$ emission is also prominent (e.g. see the
H$\alpha$ map in Kamphuis et al. 2007). The dust heating is likely
related to in-disk star forming regions at these locations.

\subsection{Major Axis Profile}

\begin{figure}
  \includegraphics[scale=0.6,angle=-90]{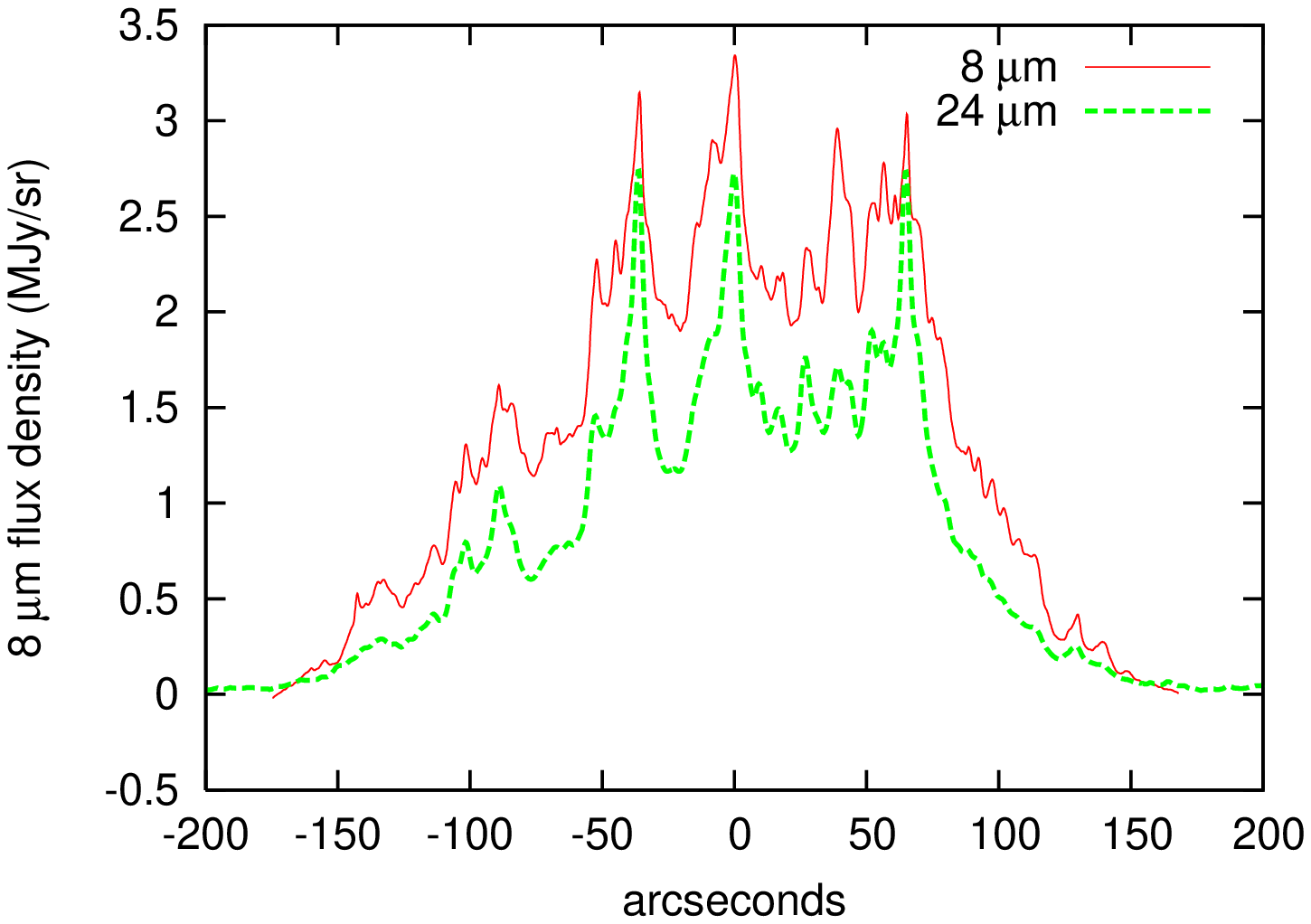}
  \caption{Surface brightness profiles along NGC 891's disk for the
$\lambda\,$8 $\mu$m band (rms noise of
0.025 MJy sr$^{-1}$)
and the $\lambda\,$24 $\mu$m band, obtained from an
average of $\pm$ 150 arcsec in $z$ above and below the plane of the
image in Fig.~\ref{fig:pah_cont}. 
Positive
along the x axis corresponds to the northern side of the disk and
negative the southern side.  
The $\lambda\,$24 $\mu$m flux
has been
normalized to the $\lambda\,$8 $\mu$m data and can be 
can be recovered by multiplying by 333. The 
$\lambda\,$24 $\mu$m noise, prior to normalization, is
0.86 MJy sr$^{-1}$.}
  \label{fig:77major_axis}
\end{figure}

On large scales, 
Swaters et al. (1997) found a north-south asymmetry 
in the HI emission, such that the southern emission is more extended radially
along the major axis.  
Popescu \& Tuffs (2003) subsequently found a similar result in the
$\lambda\,$200 $\mu$m
emission.  Asymmetries have also been
seen in the molecular distribution (see Sofue \& Nakai 1993). Given
the correlations between bands indicated above, we investigate
whether any global asymmetries are present in our data using a
representative PAH band which we take to be the Spitzer IRAC
$\lambda\,$8 $\mu$m data (Figure~\ref{fig:pah_cont}), and the
Spitzer MIPS $\lambda\,$24 $\mu$m data (Figure~\ref{fig:dust_maps}) 
since the entire galaxy has
been imaged in these wavebands (as opposed to the narrower band but
truncated ISO images of Figure~\ref{fig:pah_cont}).  The result is
shown in Figure~\ref{fig:77major_axis}.  Since the noise is very low
(see figure caption), virtually all the structure seen in this
profile is real.  The broadscale distribution of emission is
consistent with what has been seen at other MIR wavelengths (e.g.
Dupac et al. 2003) and this PAH profile also compares well with the
lower spatial resolution ISOPHOT observations (Mattila et al. 1999,
their Figure 5a).  However, to our knowledge,
Figure~\ref{fig:77major_axis} presents the highest resolution and
highest S/N major axis profile yet obtained.

The profile shows a peak at the centre and two broad peaks on either
side which, together form a
  broad 200 arcsec central plateau of PAH emission.
This emission contains the most active part of the disk encompassing
the nucleus and the secondary peaks at -40 arcsec (south of the
nucleus) and +40 arcsec to +60 arcsec (north of the nucleus). Much
substructure is also visible.
 As for asymmetries,
we can say that there is some enhancement on the northern side
within a radius of $\approx\,$ 150 arcsec and a minor enhancement
towards the south at radii larger than this.

\subsection{The Vertical Distribution \&  Extraplanar Emission}
\label{sec:minor_axis}

The `edges' of
the emission shown in Figures \ref{fig:pah_cont} -
\ref{fig:other_cont} and \ref{fig:Apah} - \ref{fig:Awide} are not smooth and show varying widths as a function
of location in the disk.  Discrete extensions, such as seen in other
galaxies (e.g. Irwin et al. 2007)
or in the DIG of this galaxy (Rossa $\&$ Dettmar 2003, see Sect.~\ref{sec:n891})
are not prominent
at the S/N of these data, although there are hints of such
discrete features at slightly lower
S/N. We know that vertical discrete features are indeed present in the
dust distribution from the optical absorption observations of
 Howk \& Savage (1997) and the H$\alpha$ observations of Rossa et al. (2004).

\begin{figure*}
  \includegraphics[scale=0.55, angle = -90]{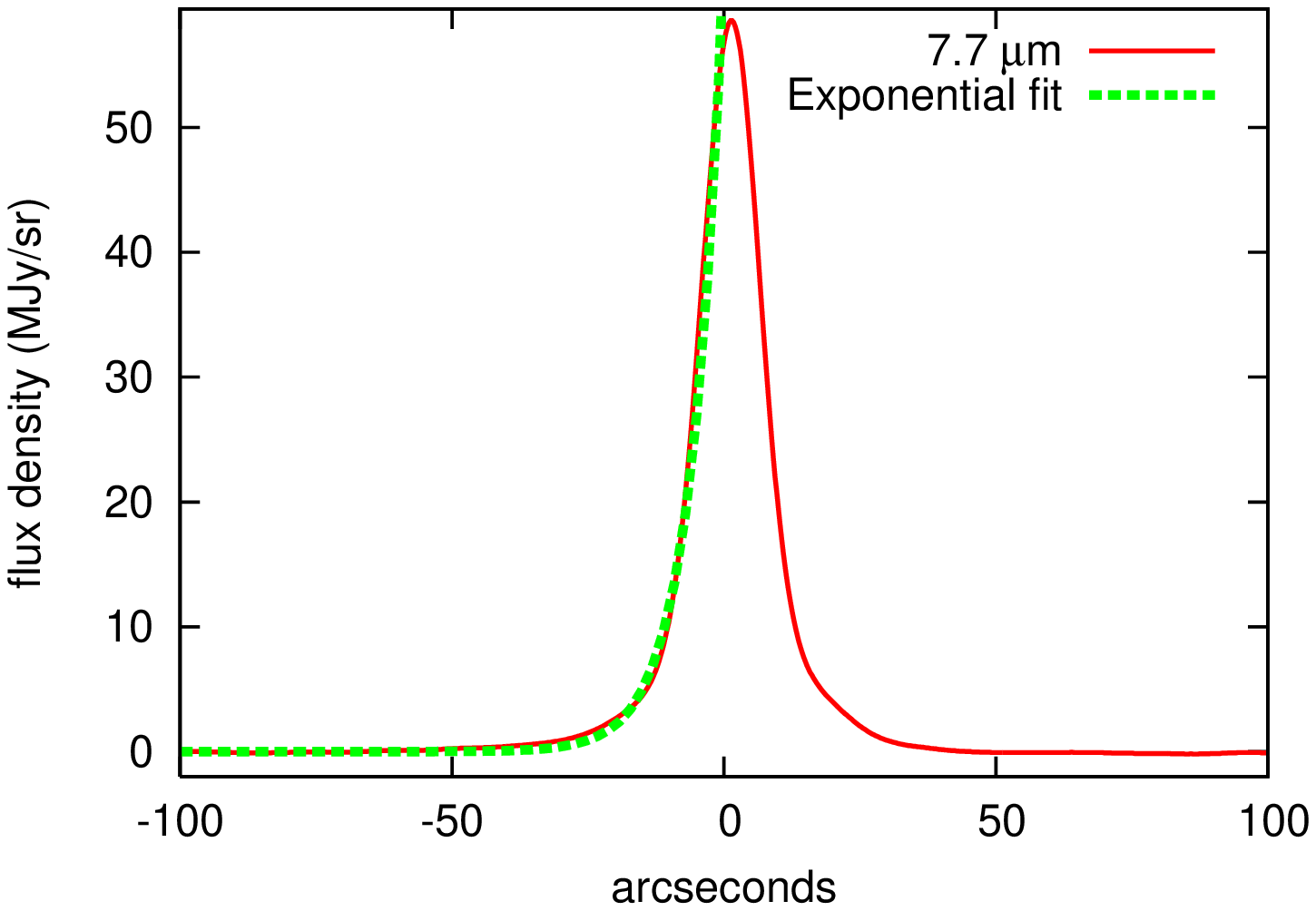}
  \includegraphics[scale=0.55, angle = -90]{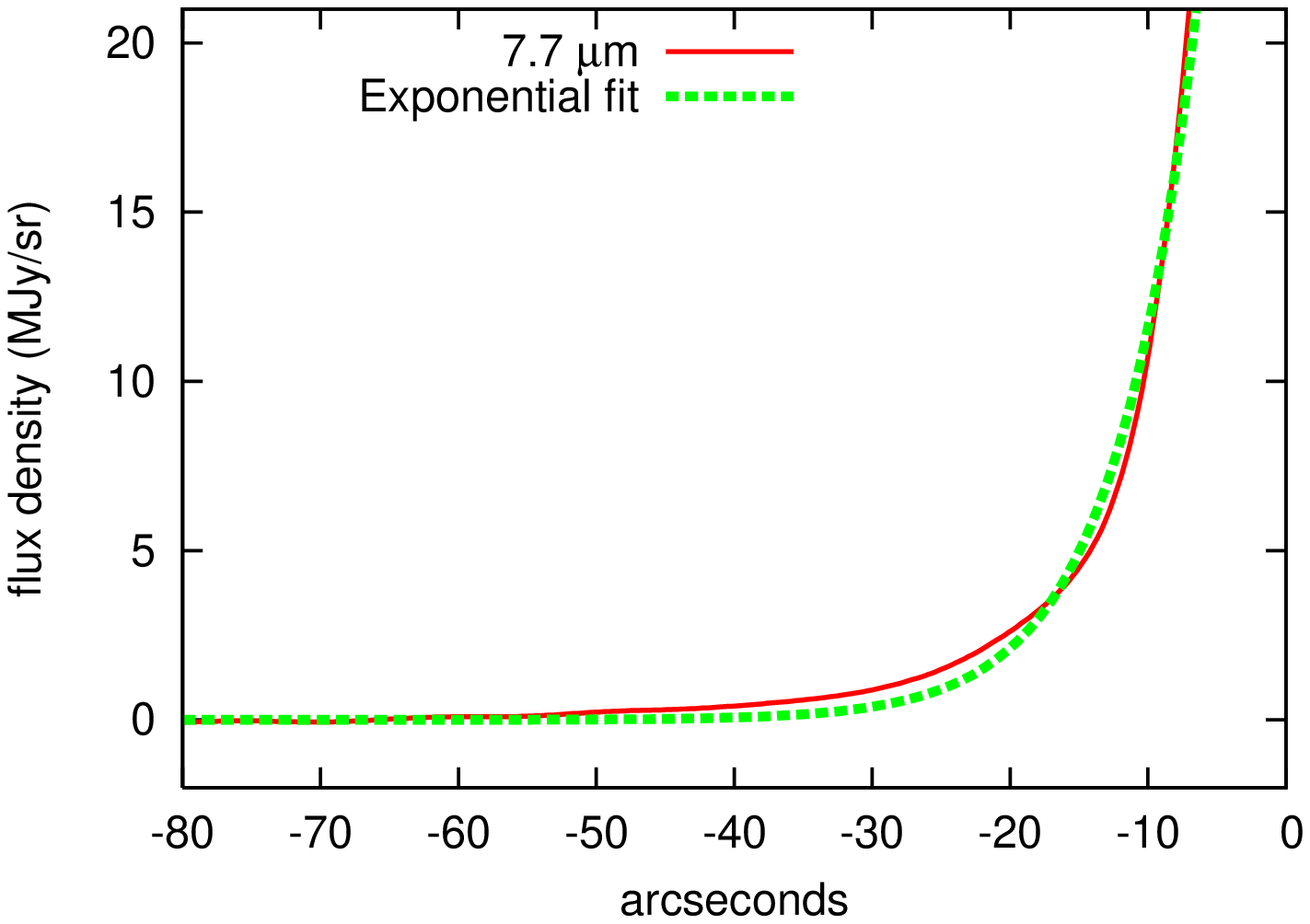} \\
  \caption{Surface brightness profile (left) for the
$\lambda\,$7.7 $\mu$m wave band (red) as a function of z
  height, with an exponential fit (green). We have zoomed into the wing region
(right) to better see the fit. The spatial resolution of the data is
7.9 arcsec.  The error bars have been omitted for clarity (see text). The vertical errors are
$\pm$ 0.028 MJy sr$^{-1}$.}
  \label{fig:77minor_axis}
\end{figure*}

\begin{figure}
  \centering
   \includegraphics[scale=0.55,angle=-90]{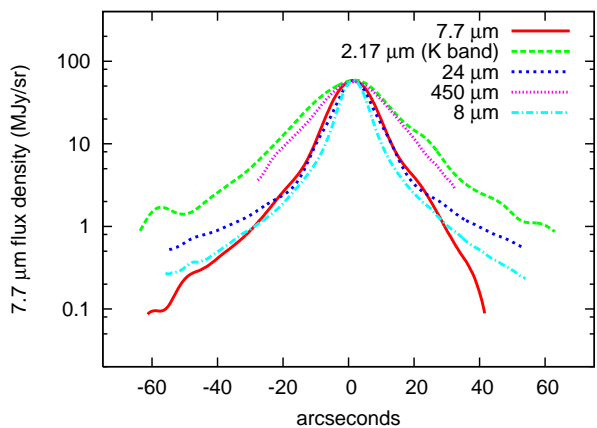}
  \caption{Surface brightness profiles for the $\lambda\,$7.7 $\mu$m
  (red, as in Figure~\ref{fig:77minor_axis}), 2MASS K band (green),
$\lambda\,$24 $\mu$m MIPS band (blue), $\lambda\,$450 $\mu$m SCUBA
band (magenta), and $\lambda\,$8 $\mu$m IRAC band (cyan)
 as a function of z height; all data have been smoothed to 7.9 arcsec resolution
 prior to forming the profile. All data have been cut off at the
 3$\sigma$ level which, prior to scaling, is 0.085, 0.033,
1.6, 2.6, and 0.13 MJy sr$^{-1}$ at $\lambda\,$7.7
$\mu$m, K band, $\lambda\,$24 $\mu$m, $\lambda\,$450 $\mu$m, and
$\lambda\,$8 $\mu$m, respectively.  The horizontal error bars are $\pm$ 4 arcsec.}
  \label{fig:77onk450_24}
\end{figure}

\begin{figure}
  \centering
  \includegraphics[scale=0.6,angle=-90]{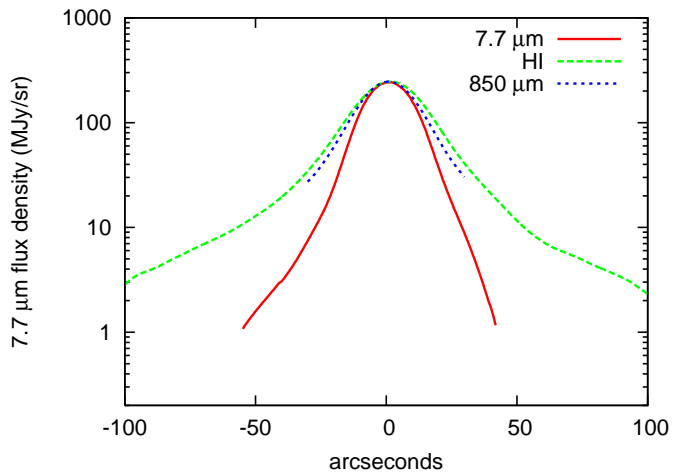}
  \caption{Surface brightness profiles for $\lambda\,$7.7 $\mu$m (red),
  H I emission from Swaters (1997) (green)
  and $\lambda\,$850 $\mu$m SCUBA band (blue) as a function of z height. These
  data have been smoothed to 20 arcsec resolution and have been cut off at the
 3$\sigma$ level (with the exception of the HI). The horizontal error bars are
$\pm$ 10 arcsec.}
  \label{fig:77onHI850}
\end{figure}

To further investigate the presence of extraplanar dust in these
data, we have averaged the emission parallel to the major axis (within
$\pm$ 2.5 arcmin radially from the minor axis)
to create an average vertical profile.  While such averaging removes
information about any discrete features, it does increase the S/N,
improving the possibility of detecting widespread high latitude
dust.  We have also taken a subset of the data sets and smoothed
them to a common spatial resolution (either 7.9 arcsec or 20 arcsec)
so that we can compare the vertical extent of the emission in
various bands. Results are shown in Figure~\ref{fig:77minor_axis},
in which a typical profile (in this case the narrow band
$\lambda\,$7.7 $\mu$m ISO PAH band) is shown in linear scale
to emphasize the exponential fit, and in a log scale in
Figures~\ref{fig:77onk450_24} and \ref{fig:77onHI850}
which compare
the various profiles at 7.9 arcsec and at 20 arcsec resolution,
respectively.  As was the case for the major axis profile, these
average $z$ slices have very small vertical error bars (see
Figure captions), and have been omitted from the figures for clarity.
 The main error, for the purpose of comparing
the width of the emission in the various wavebands, is
in the horizontal direction, associated with the spatial resolution
($\pm$ 4 arcsec for Figures \ref{fig:77minor_axis} and
~\ref{fig:77onk450_24},
and $\pm$ 10 arcsec for Figure
\ref{fig:77onHI850}).

Figure~\ref{fig:77minor_axis} is a typical MIR vertical profile and
shows that an exponential fall-off of emission fits the data well in
both the core and wing region.  This is also true of the other MIR
bands (fits not shown, but see also the exponential fits of Burgdorf et
al. 2007 and Rand et al. 2008).
We have illustrated the narrow band $\lambda\,$7.7 $\mu$m
ISO profile in particular, since it is excellent at isolating PAH
emission. Note that the wider waveband $\lambda\,$8 $\mu$m IRAC profile
matches the $\lambda$7.7 $\mu$m ISO profile to within errors
 (cf. Figure~\ref{fig:77onk450_24}).  Single-component exponentials
fit all data sets well, although some departures are seen at very low levels. 
The exponential scale heights that result
from these fits, $z_e$,
are not sufficiently different from the spatial resolution to be convincingly
presented.
However, of the dust-related
data sets at their original un-smoothed resolution,
the IRAC $\lambda\,5.8$ $\mu$m and IRAC $\lambda\,8.0$ $\mu$m data sets have
high enough spatial resolution (Table~\ref{tab:obs_spitzer}) 
for a $z_e$ measurement.  For these, we find (over the same 5 arcmin averaged region
as Figure~\ref{fig:77minor_axis})
$z_e\,=\,302\,\pm\,70$ pc and
$z_e\,=\,248\,\pm\,50$ pc 
at $\lambda\,5.8$ and $8.0$ $\mu$m, respectively, where the error bars are dominated
by the variation between the north-west and south-east sides of the galaxy.
Our scale height for the 
PAH-dominated $\lambda\,8.0$ $\mu$m band over this averaged region
 is slightly lower than the
single-component $z_e$ estimate of 330 - 530 pc
found for $\lambda\,17~\mu$m PAH emission measured at a single position 
by Rand et al. (2008).  Given the possible effects of unknown in-disk
extinction on the vertical scale height (see Rand et al. 2008) as well as
uncertainties related to aperture corrections for the IRAC data 
(Sect.~\ref{sec:MIPS_obs}), two-component fits are not warranted.

Figure~\ref{fig:77onk450_24} shows a comparison between
bands at high spatial resolution, for which the error in $z$ is
$\pm\,$ 4 arcsec, or 0.2 kpc. 
The logarithmic scale emphasizes lower levels for which there are
some departures from perfect exponentials. The broadest component is clearly in
K band, due to the dominance of starlight in this band and the
presence of the galaxy's bulge.  The cool dust component, illustrated by the
$\lambda\,$450 $\mu$m emission, is the widest dust band and is
measured to $z\,\approx\,$29 arcsec (1.3 kpc) at 3$\sigma$
(average $z$ of both sides of the vertical profile). The PAH
emission ($\lambda\,$7.7 and 8 $\mu$m) and warm dust emission
($\lambda\,$24 $\mu$m) have $z$ widths that agree within errors.
Also, their measurable maximum extents are to $z\,=\,$ 49 arcsec
(2.3 kpc), 53 arcsec (2.5 kpc), and 54 arcsec (2.5 kpc) at
$\lambda\,$7.7, $\lambda\,$8 $\mu$m, and $\lambda\,$24 $\mu$m,
respectively, again agreeing within errors. 
  Although the maximum extent of the various components
depends on the varying S/N, it is clear that
each of these components reveal emission at $z$ in excess of
1 kpc. 
It is therefore
 clear that NGC~891 has extraplanar emission in warm dust, cool dust, and
PAHs.  Kamphuis et al. (2007) have also recently examined the MIPS
$\lambda\,$24 $\mu$m emission from NGC~891 and find emission to 
$z$ of 2.3 kpc, in agreement with our results.  Burgdorf et
al. (2007) have also found extraplanar dust emission extending to
 $z\,=\,$5 kpc, at a specific location rather than an average,
and at the wavelengths, $\lambda\,$16 and $\lambda\,$22 $\mu$m.

As for the lower resolution profiles (Figure~\ref{fig:77onHI850}),
the spatial resolution (20 arcsec = 0.9 kpc) is less useful for
examining extraplanar gas quantitatively.  However, 
it is apparent that the H I distribution is widest (see also
Sect.~\ref{sec:intro}), followed by cold dust at $\lambda\,$850
$\mu$m and PAHs.


\section{Spectral Energy Distribution}
\label{sec:sed}

\subsection{Constructing the SED}
\label{sec:sed_construct}

We have constructed spectra using the flux from different regions of
the galaxy in each of the wave bands. The adopted boxed regions are
shown in Figure \ref{fig:n891_wboxes} and described in Table
\ref{tab:n891regions}. They represent a potential diversity of
environments within which the S/N of the data was sufficient at all
wavebands for good results. Each box is 18 arcsec $\times$ 18 arcsec
(838 pc $\times$ 838 pc) in size, a value chosen because all images
were smoothed to 18 arcsec resolution to match the lowest resolution
 $\lambda\,$70 $\mu$m data before the measurements were made.
The four in-plane boxes represent the centre of the galaxy and three
disk regions. The centre contains the disk that passes along our
line of sight (old and young stars and ISM), old stars from the
bulge and the nucleus.
The disk region in the southern half is located at a local maximum.
Although there may be some contributions from the bulge to the
northern and southern disk regions, our results for these regions should
be strongly dominated by the disk components that are along the line of sight.
All in-disk regions fall within the broad plateau of emission shown in
Figure~\ref{fig:77major_axis} including the outer
 north disk at a radial distance of
96 arcsec (4.47 kpc). Finally there are two halo regions, each
centered between 1 and 2 kpc above the plane. 
These regions should therefore provide one of the first MIR spectra of 
halo ISM in an external galaxy (with Rand et al. 2008 being the only other
one at the time of this publication).


For the flux measurements, the error bars are due to the published
flux errors given for each instrument. To summarize from
Sect.~\ref{sec:obs_datared}, these are approximately 20$\%$ for
ISOCAM, 5$\%$ for IRAC, 4$\%$ - 15\% for MIPS, 2$\%$ for 2MASS, and
15 - 25$\%$ for SCUBA. These are the errors that dominate when
comparing between different wave bands. We present the spectra
fitted by SED models in Figure \ref{fig:all_seds}. Note that we only
use upper limits for the 70 micron measurements of the halo because
the extraplanar region is severely affected by latent image effects
that create streaking in the images. This data point is carrying
lower weight in the SED modelling. The next section describes this
model.

\begin{table}
\begin{minipage}{85mm}
  \caption{Regions of NGC 891 Chosen for Spectral Analysis}
  \label{tab:n891regions}
  \begin{tabular}{llll}
    \hline
    & name of & RA $\&$ DEC\footnote{Coordinates refer to J2000.0} & offset (R, $z$)\footnote{Distance
from the center of the galaxy along and perpendicular to the major
axis, respectively.} \\
    $\#$ & region & (h m s), ($^{\circ}$ $^{\prime}$ $^{\prime\prime}$)
	& (kpc, kpc) \\
    \hline
    1 & centre  & 2 22 33.092, 42 20 50.26 & 0, 0\\
    2 & south disk & 2 22 30.928, 42 20 05.26 & 2.51, 0\\
    3 & inner north disk & 2 22 35.258, 42 21 52.26 & 2.79, 0 \\
    4 & outer north disk & 2 22 37.153, 42 22 30.26 & 4.47, 0\\
    5 & central halo & 2 22 30.927, 42 21 06.26 & 0, 1.40\\
    6 & halo over disk & 2 22 33.002, 42 21 59.26 & 2.79, 1.26\\
    \hline
  \end{tabular}
\end{minipage}
\end{table}

\subsection{Modeling the SED}
\label{sec:model}

The model we use fits the observed SEDs for a situation in which a
variety of dust components with realistic optical properties are
exposed to an incident radiation field from a mixture of stars, all
within the same beam.
The model includes the following IR emission sources: \\
1) PAHs - which can be singly ionized ($PAH^+$) or neutral ($PAH^{\circ}$), \\
2) Dust - which include VSGs and large grains composed of graphites and silicates, and \\
3) Old Stars

We adopt the dust size distributions derived by Zubko et al.\ (2004)
by fitting the diffuse Galactic ISM emission, extinction and
depletion patterns, with PAH, graphite and silicate grains, and
assuming solar abundances. Thus we are assuming that the instrinsic
dust properties in NGC 891 are similar to the ones in the MW. In
fact, this is a good assumption because both galaxies are solar
metallicity spiral galaxies (12+log(O/H)=8.9, Otte et al. 2001). The
PAHs are assumed to have the optical properties (e.g. absorption
cross sections) from Draine $\&$ Li (2007a). Note that, for the
purpose of the model, PAHs will be included in the `dust' component,
rather than considered as `large molecules', given the
similarity between the largest PAHs and smallest carbon grains. Thus
the model parameter, $M_d$, represents the \textit{total} dust mass;
this includes PAHs, graphites and silicates.

The PAH component introduces two parameters: the mass ratio of PAHs
to total dust normalized to the Galactic value of 0.046 ($f_{PAH}$):
\begin{equation}
f_{PAH}= \frac{1}{0.046} \frac{M_{PAH^{\circ}} + M_{PAH^+}}{M_d}
\label{eqn:pah_dust_ratio}
\end{equation}
and the ratio of ionized PAHs to total PAHs ($f_{PAH^+}$):
\begin{equation}
f_{PAH^+} = \frac{M_{PAH^+}}{M_{PAH^{\circ}} + M_{PAH^+}}
\label{eqn:ionized_pah}
\end{equation}
where $M_{PAH^{\circ}}$ and $M_{PAH^+}$ are the masses of
the neutral and ionized PAHs respectively,
and $M_{PAH}$ = $M_{PAH^{\circ}}$ + $M_{PAH^+}$.

The stellar continuum makes a large contribution to the SED at NIR
wavelengths. It needs, as a free parameter, the total stellar mass,
$M_{oldstar}$. $M_{oldstar}$ is used to compute the stellar
emission, $L_{\nu}(\lambda)^*$\footnote{The subscripts $\nu$ or
$\lambda$ indicate that this is a monochromatic term, and the
$\lambda$ in brackets indicates a wavelength dependence.}.
 Note,
however, that the resulting value for
$M_{oldstar}$ is not actually a proper
measurement of total stellar mass.  The latter would require data
 at shorter wavelengths,
 different extinction geometries, varying the stellar age and eventually
the Initial Mass Function (IMF).  Rather, $M_{oldstar}$ is a
parameter that is varied to understand the contribution from evolved
stars into the MIR.

The dust must be excited by optical-UV radiation to emit. We adopt
the shape of the interstellar radiation field (ISRF) of the Galactic
diffuse ISM by Mathis et al. (1983), $I_{\lambda}^*(\lambda)$.

The adopted ISRF, \textit{U}, is normalized by the local Solar
neighbourhood value (taken to be 2.2 $\times$ 10$^{-5}$ W m$^{-2}$):
\begin{equation}
  U \equiv \frac{\displaystyle \int_{0.09 \mu m}^{8 \mu m} I_{\lambda}^*( \lambda ) \,d{\lambda}}{2.2 \times 10^{-5}}
\label{eqn:U}
\end{equation}
where the lower limit on the integral is set at the Lyman break and
the upper limit approximates the value at which excitation is no
longer significant.

The monochromatic luminosity of dust exposed to \textit{U} per unit
dust mass, $l_{\nu}(U,\lambda)$, is:
\begin{equation}
  l_{\nu}(U,\lambda) \equiv \frac{dL_\nu (U,\lambda)}{dM_d (U)}
\label{eqn:lum_diff}
\end{equation}
This monochromatic luminosity can be expressed
as a 
sum of the three dust components
(PAHs, silicate and graphite):
\begin{eqnarray}
  l_{\nu}(U,\lambda)=f_{PAH}\biggl[f_{PAH^+}l_{\nu}^{PAH^+} (U,\lambda) + (1-f_{PAH^+}) \nonumber \\
  \times l_{\nu}^{PAH^{\circ}}(U,\lambda)\biggr] +
  l_{\nu}^{sil.}(U,\lambda) + l_{\nu}^{gra.}(U,\lambda)
\label{eqn:luminosity}
\end{eqnarray}
Here $l_{\nu}^N$(U,$\lambda$), where N = $PAH^+$, $PAH^{\circ}$,
sil. and gra., are the luminosities of the ionized PAHs, neutral
PAHs, silicates, and graphites respectively.

With an admixture of dust components in an interstellar
 radiation field, we then need to
consider the possible range of environments within which these
components could be found.  Following Dale et al. (2001), this can
be done via the parameter, $\alpha$, which relates the heated dust
mass, $M_d$, to the radiation field,
\textit{U}: 
\begin{equation}
  dM_d(U) \propto U^{-\alpha}dU
\label{eqn:Md_U}
\end{equation}
between $U_{min}$ and $U_{max}$.
The range of
$\alpha$, which is defined by Eqn.~\ref{eqn:Md_U} is typically
1 to 2.5,
depending on whether the dust is in a diffuse medium
($\alpha$ $\approx$ 2.5) or whether the dust is in a dense medium
($\alpha$ $\approx$ 1). Since the area measured contains
contributions from both kinds of media, the $\alpha$ from the
model is a characteristic value for the measured region and will
likely fall somewhere between 1 and 2.5.

Therefore, with Equations (\ref{eqn:lum_diff}) and (\ref{eqn:Md_U}),
and the stellar component added, the total SED can be expressed as
\begin{eqnarray}
  L_\nu(\lambda) = \biggl[ \frac{(1 - \alpha)M_d}{U_{max}^{1-\alpha} - U_{min}^{1-\alpha}} \int_{U_{min}}^{U_{max}} l_\nu(U,\lambda)U^{-\alpha}
    \,dU + L_\nu^*(\lambda) \biggr] \nonumber \\
    \times \exp[-\tau(\lambda)]
\label{eqn:L_tot}
\end{eqnarray}
where the optical depth, $\tau$ = A$_V$/1.086, (and A$_V$ is the
visual extinction), and \textit{L}*($\lambda$) comes from Fioc $\&$
Rocca-Volmerange's (1997) model\footnote{The NIR old stellar
continuum is roughly independent of the age of the populations for a
burst $>$ 1 Gyr. The stellar continuum is modeled with the
population synthesis code \url{PEGASE}.} for a burst of 5 Gyr and a
mass of stars, $M_{oldstar}$. The $\exp\left[-\tau(\lambda)\right]$
term accounts for the extinction in the `slab' geometry that we have
adopted.  Note that the visual extinction, A$_V$, is constrained by
the J, H, and K band ratios.  A `true' value of A$_V$ requires additional
measurements at visible wavelengths, so this extinction should be
deemed `representative', rather than exact, and useful for comparing
different regions in the galaxy.

To summarize, the eight parameters the model is fitting are:
\begin{itemize}
\item The mass of dust ($M_d$) in units of $M_{\odot}$,
\item The ratio of mass of PAHs to mass of dust normalized to the Galactic value (0.046) ($f_{PAH}$),
\item The ratio of mass of ionized PAHs to total PAHs ($f_{PAH^+}$),
\item The mass of the old stellar population ($M_{oldstar}$) in units of $M_{\odot}$,
\item Minimum and maximum heating intensities ($U_{min}$ $\&$ $U_{max}$
respectively) normalized to the local radiation environment (2.2 $\times$ $10^{-5}$ W m$^{-2}$),
\item The visual extinction (A$_V$), and
\item $\alpha$ defined in Equation \ref{eqn:Md_U}.
\end{itemize}

Finally, we note that the parameters $\alpha$, $U_{min}$ and
$U_{max}$ do not have a strong physical meaning when taken
independently. However, these parameters together quantify the
distribution of dust mass through different environments. The
fraction $f_{\rm cold}$ of dust mass exposed to radiation densities
lower than $U\lesssim8.5$ (corresponding to a silicate equilibrium
temperature of 25~K) is:
\begin{equation}
  f_{\rm cold} = \int^{U(25\;\rm K)}_{U_{min}} \frac{(1-\alpha)\,U^{-\alpha}}{U_{max}^{1-\alpha} - U_{min}^{1-\alpha}} \,dU
\label{eqn:temp_frac}
\end{equation}
These quantities will also be calculated from the model.

This model is integrated within each filter, with the proper flux
convention, and then compared to the observations, by minimizing the $ 
\chi^2$. In addition, we estimate the error on the value of the parameters as  
follows. For each observed SED, $\displaystyle\left\{F_\nu(\lambda_i)\right\}_i 
$, with error bars $\displaystyle\left\{\Delta F_\nu(\lambda_i)\right\}_i$, we
perform several fits of $\displaystyle\left\{F_\nu(\lambda_i)+\theta_i
\times\Delta F_\nu(\lambda_i)\right\}_i$, where $\theta_i$ are  
independant normally distributed random variables.
Therefore, such a fit provides the values of the parameters for one  
particular realization of the error bars.
By iterating over several $\theta_i$ sets, we obtain the distribution of  
each parameter value. The dispersion of this distribution provides the error bar 
on the parameter.


\subsection{SED results for NGC 891}
\label{sec:sed_results}

\begin{figure*}
  \includegraphics[scale=0.49]{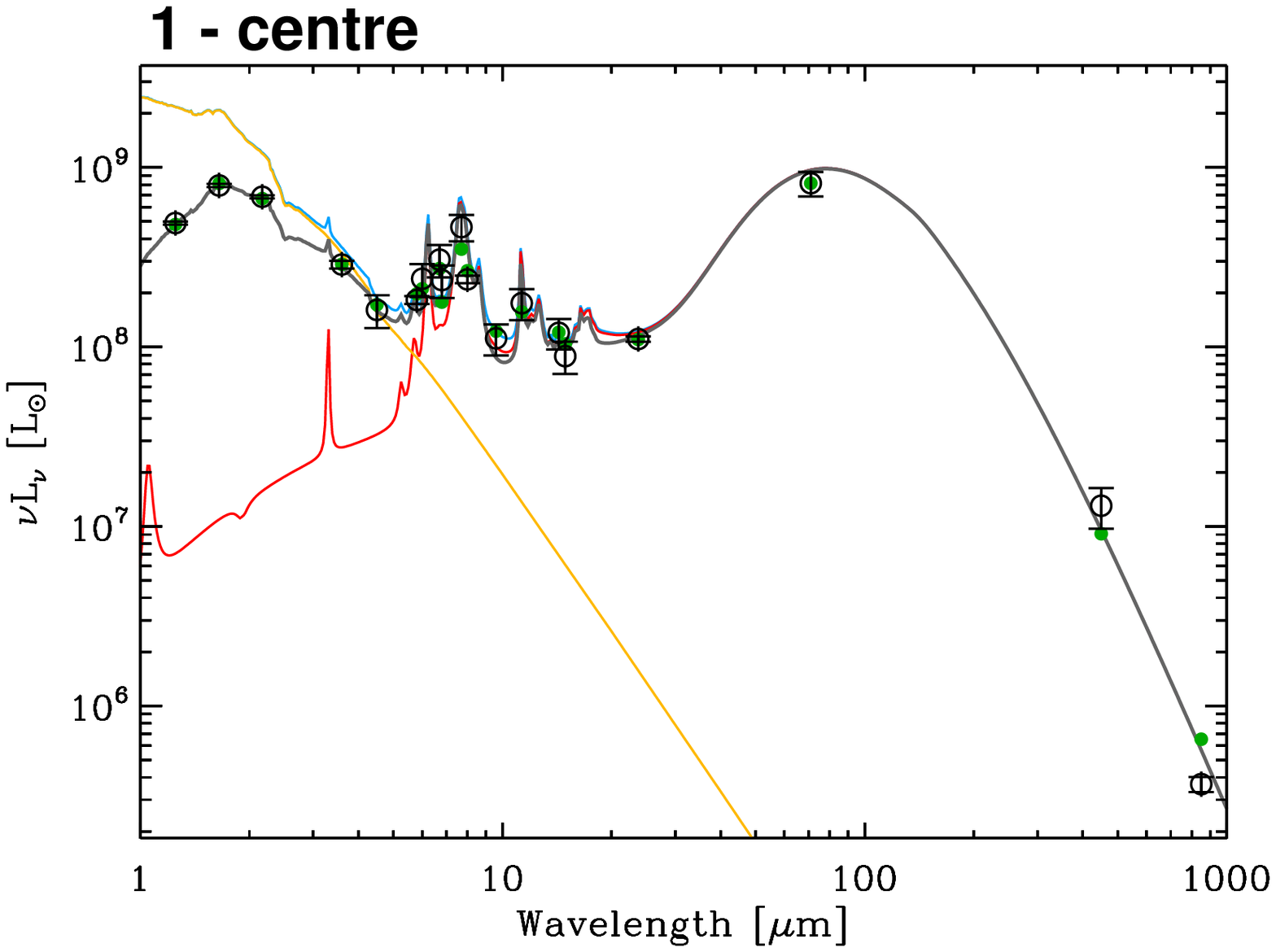}
  \includegraphics[scale=0.49]{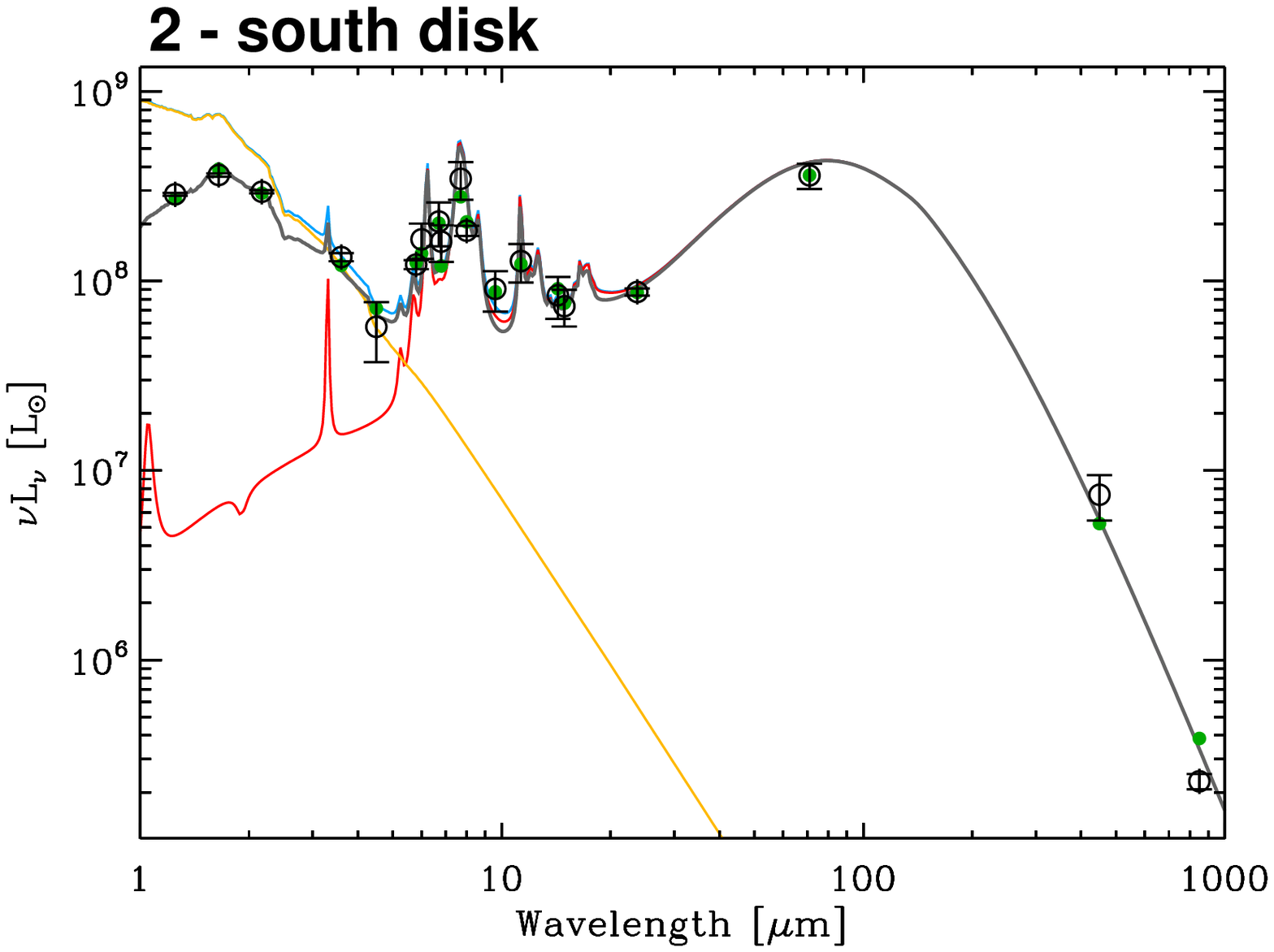}
  \includegraphics[scale=0.49]{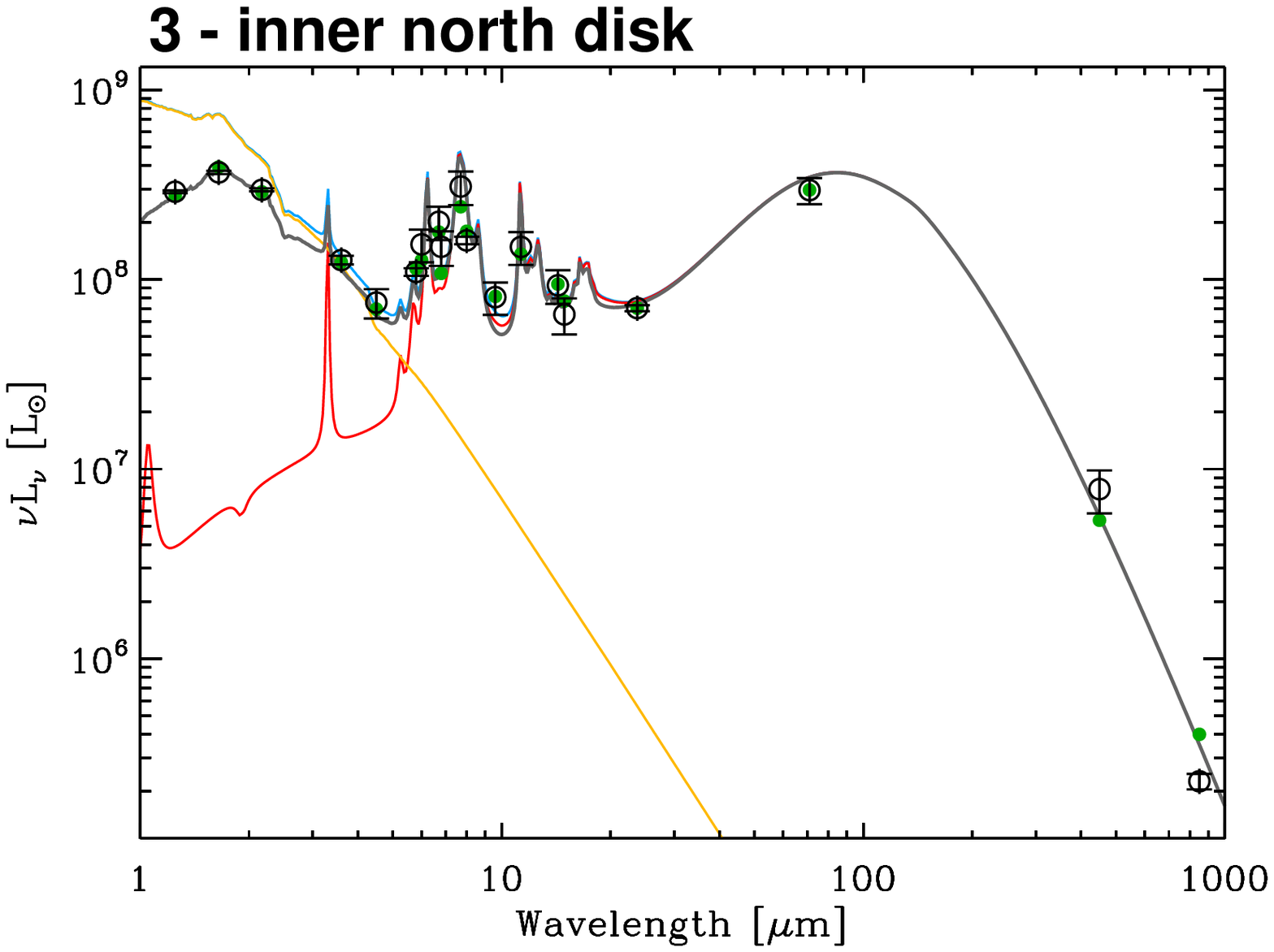}
  \includegraphics[scale=0.49]{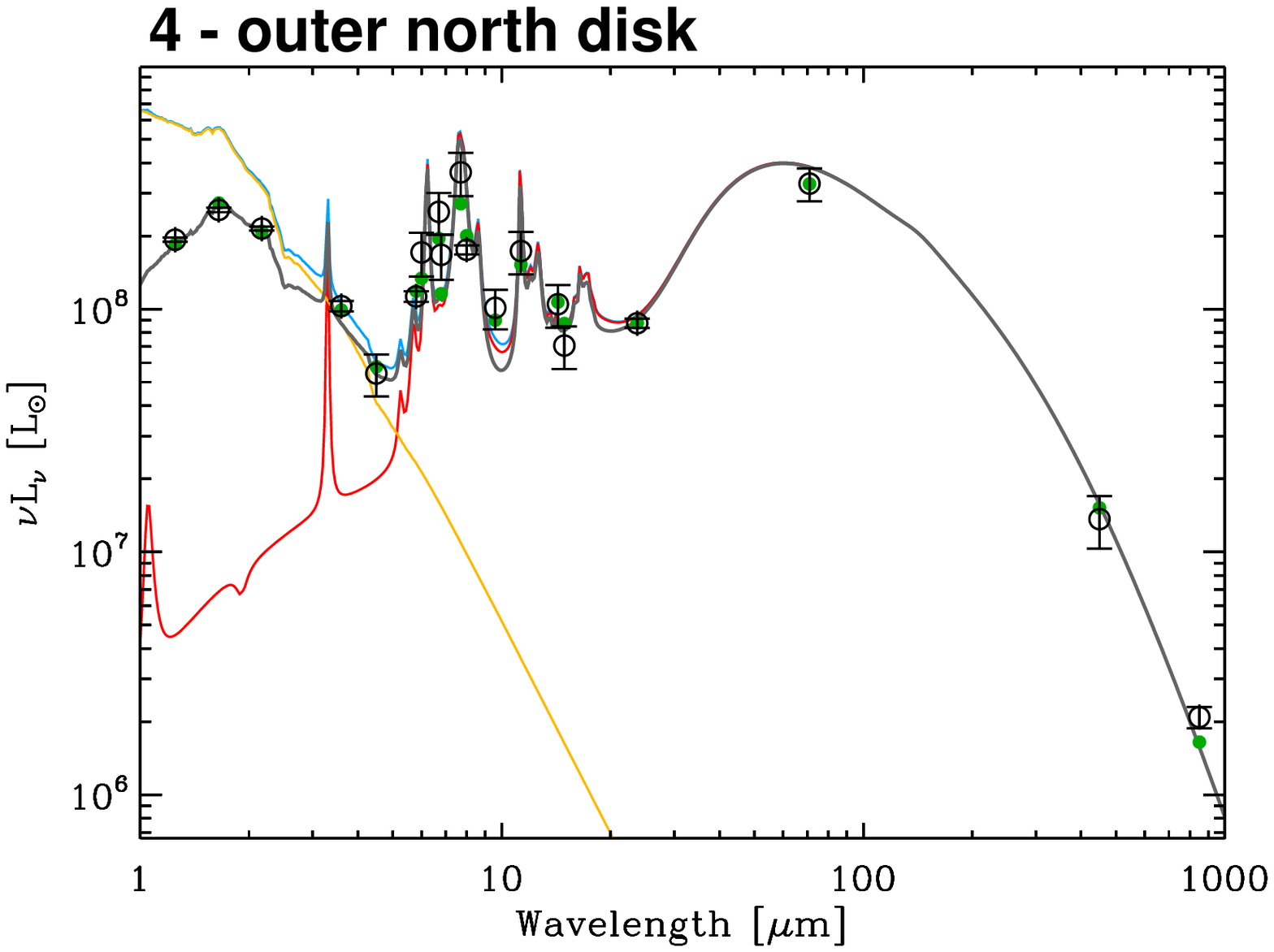}
 \includegraphics[scale=0.49]{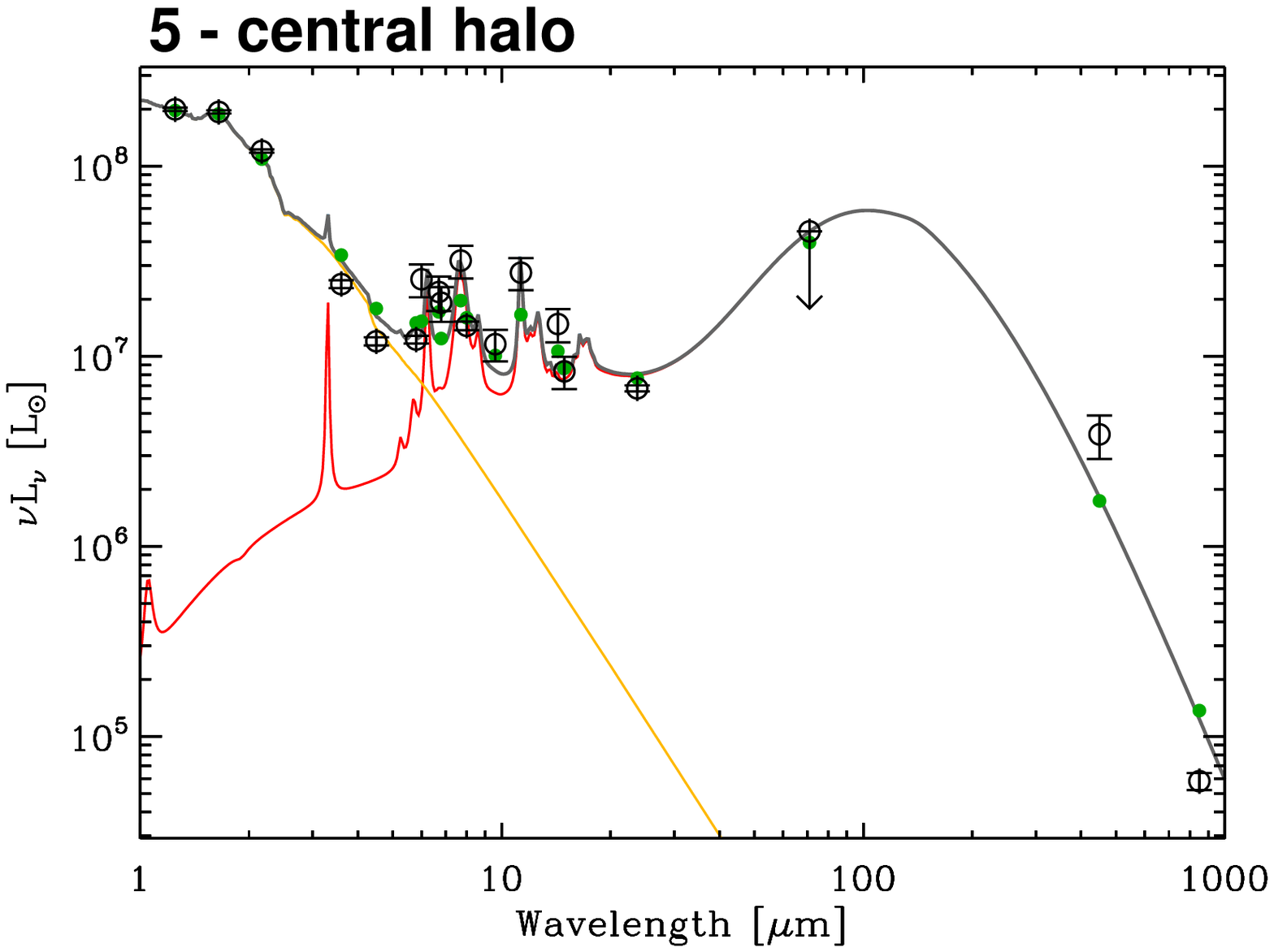}
 \includegraphics[scale=0.49]{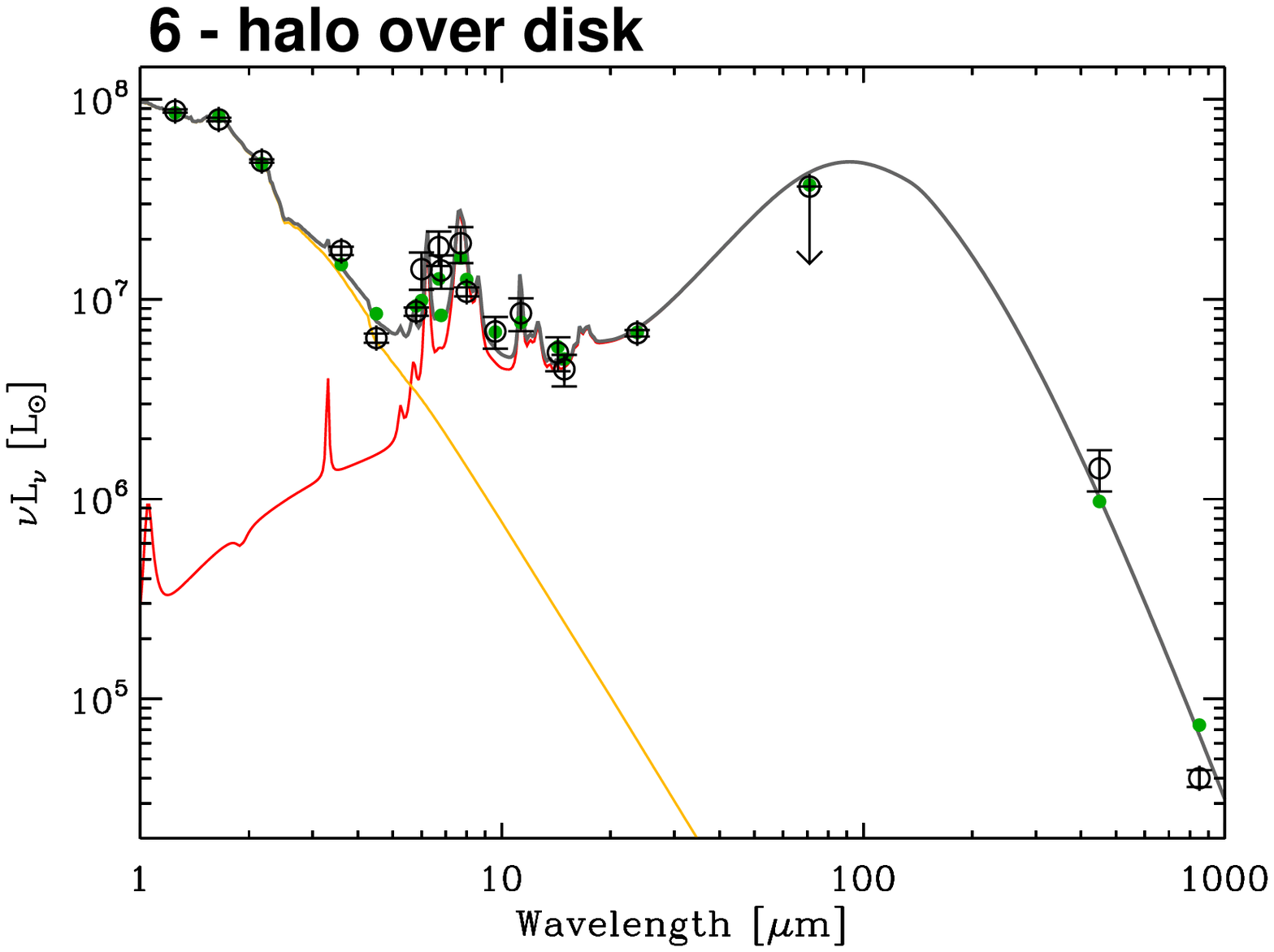}
  \caption{SEDs for the various regions
of NGC~891 given in Table~\ref{tab:n891regions}, fit
  with the model described
  in Sect. \ref{sec:sed}. The yellow line represents the stellar contribution
corrected for extinction, the red line
  represents the dust and PAH contribution, the grey line is the total line
of best fit and the blue line
  is the SED without extinction. The data points are
circles with error bars and the green points are
  the model results in the observed bands.}
  \label{fig:all_seds}
\end{figure*}

Figure \ref{fig:all_seds} shows the observed SEDs for
the six measured regions in NGC 891 as well as the model fits.
The yellow line represents the stellar contribution
corrected for extinction, the red line represents the dust and PAH contribution, 
the grey line is the total line of best fit and the blue line is the SED without 
extinction. The data points are circles with error bars and the green points are
the model results in the observed bands.

In general, the model results agree with the data to within estimated errors
throughout the NIR to the sub-mm.  The exceptions are
the $\lambda\,$450 $\mu$m points, which are often too high.
One possibility is that the observed discrepancy is real and that the dust
properties in this part of the spectrum are peculiar.
 However, it is also possible
that the $\lambda\,$450 $\mu$m fluxes may be in error. We have accepted the
original calibration from previously published results at this
wavelength (Sect.~\ref{sec:other_obs}) and both Israel et al (1999)
and Alton et al. (1998), who observed this galaxy with 2 independent
sets of observations, give consistent flux values.  Nevertheless,
the total error for this point may have been underestimated.

The FIR flux densities in the models peak at shorter wavelengths (hotter)
compared to what has been observed in many other galaxies (e.g.
Dunne $\&$ Eales 2001; Bendo et al. 2003; Dale et al.
2005; Draine et al. 2007b). For example, a better constrained total SED peaks at
around $\lambda\,$100 $\mu$m (Galliano et al. 2008a), whereas many of our SEDs are peaking at 
about $\lambda\,$60 - 70 $\mu$m (Figure \ref{fig:all_seds}). 
However, in this case, it is the relatively
low $\lambda\,$850 $\mu$m fluxes that prevent the SEDs from having a more powerful cold component.
Adding additional FIR data in the 
$\lambda\,$100 - 200 $\mu$m range would tend to make the SEDs a little 
colder, but it would also make the fit of the SCUBA data even worse.
Nevertheless, given that each region has 
 data points at the same wavelengths, the SED model results for
the various regions can still be compared to
each other in a meaningful way.


The four in-disk regions all show similarly shaped SEDs which peak
in the $\lambda\,$70 to 80 $\mu$m wavelength regime, including the outer north
disk which might have been expected to show a `colder' (longer
wavelength) peak given its larger radial offset from the center
(Table~\ref{tab:n891regions}). However, as indicated in
Section~\ref{sec:sed_construct}, the position of the outer north
disk point is 96 arcsec from the centre which coincides with a broad
  peak in the H$\alpha$ emission
(see image in Kamphuis et al. 2007).  It also corresponds to a peak
in the surface brightness profile along the disk (see
Figure~\ref{fig:77major_axis}). Therefore, its SED is likely more
representative of a SF region than of the far outer regions of a
typical galactic disk. The halo SEDs, on the other hand, do peak at
longer wavelengths suggesting the dominance of colder dust at high
latitudes. Adopting lower $\lambda\,$70 $\mu$m values should not change this
conclusion.


The halo SEDs differ from the disk SEDs in other significant ways.
First of all, they show an overall lower surface brightness by about
a factor of 10 (see
absolute values on ordinate axes, the apertures being indentical),
as might be expected for a faint halo region. They also reveal that
virtually no extinction (by our representative A$_V$, see 
Section~\ref{sec:model}) is required to match the NIR emission which
is dominated by starlight. Again, this is not surprising, given the
contrast between observing stellar emission through a dense 
dusty, edge-on disk, or observing starlight in a faint halo region
with mainly diffuse tenuous dust.

The other contrast between the halo and disk SEDs are in the
relative brightness of the MIR in comparison to other parts of the
spectrum. The halo SEDs show a lower brightness in the MIR, where
PAHs are the dominant contributor, in comparison to disk SEDs. For
example, the three in-disk regions (not including centre) all show
similar ratios of PAHs to old stars, as measured by the
$\lambda\,$7.7 $\mu$m peak to the dust-corrected $\lambda\,$1 $\mu$m
peak ratio, ranging from $\approx$ 0.6 to 0.9. However, this ratio
in the halo is somewhat depressed in comparison to the disk, with
values in the range
 $\approx$ 0.2 to 0.4.  As we showed in Figure~\ref{fig:77onk450_24},
the stellar halo, as measured by the K-band flux, is also broader
than the $\lambda\,$7.7 $\mu$m halo emission.  Therefore, the lower
PAH/stellar ratio in the halo is consistent with the vertical distribution
seen earlier.

PAHs in the halo are also depressed in comparison to cold dust. The
three in-disk regions (again, not including the central region) show
similar PAH/cold dust ratios, as measured
 by the
$\lambda\,$7.7 $\mu$m peak to the FIR peak ratio.
We cannot compare this ratio
directly for the halo SEDs since the latter do not have well defined peaks,
 with
only upper limits at $\lambda\,$70 $\mu$m.  Using the $\lambda\,$7.7
$\mu$m/$\lambda\,$850 $\mu$m ratio instead to probe the PAH/cold
dust emission, we find that the disk SEDs have significantly higher
ratios, of order 2000, whereas the halo ratios are of order 300.

Thus, there is a deficit of PAH emission in the halo
\textit{with respect to the stars and cold
dust}, as our global halo profiles from Sect. \ref{sec:minor_axis} suggested
(Figures \ref{fig:77onk450_24} and \ref{fig:77onHI850}).
In fact, the PAH strength is also lower
in the halo than in the centre of the galaxy,
though this is not quite as obvious from Figure~\ref{fig:all_seds}.
Rather, it is a result of the model fitting, the full
results of which we provide in Table \ref{tab:model_params}.

From Table \ref{tab:model_params} we find that there is no significant difference between
$M_d$, $f_{PAH^+}$, $\alpha$, $A_V$
and $M_{oldstar}$ for the disk regions, although
$M_{oldstar}$ is predictably higher for the central region.
Except for $\alpha$, these all of the parameters are also predictably lower
for the halo regions.
 However, $f_{PAH}$, $U_{min}$, and $U_{max}$, and $f_{cold}$
show significant variation, in general.


We will discuss all of these results more fully in Sects. \ref{sec:dis_disk} and
\ref{sec:dis_halo}.

From the SED model, the masses of the PAHs, $M_{PAH}$,
 for the six regions are implicitly
included in Table~\ref{tab:model_params} from
$M_{PAH}\,=\,0.046\,f_{PAH}\,M_d$.  As with the other modeled masses
given in Table~\ref{tab:model_params},
PAH masses will be specific to
the adopted size and location
 of the region.  Nevertheless, it is significant that
these quantities can be calculated from the model.
Few mass estimates for PAHs in external galaxies exist at this time.
Indeed, we are the first
to determine PAH mass estimates specifically for a halo region.
\begin{landscape}
\begin{table}
 \begin{minipage}{50cm}
 \caption{Modeling Parameter Results for Different Regions of NGC 891.}
 \label{tab:model_params}
{\small
   \begin{tabular}{lllll}
     \hline
parameter\footnote{See text Sect. \ref{sec:model} for the definition of the parameters.}\footnote{The parameters $\alpha$, $U_{Min}$
and $U_{Max}$ are correlated. That is the reason why they span a wide range of values and are not normally
distributed.} & centre (1) & south disk (2) & inner north disk (3) & outer north disk (4) \\
\hline
$M_d\;(M_\odot)$& $(7.3\pm1.9)\times10^5\;(26\%)$& $(4.9\pm1.4)\times10^5\;(28\%)$& $(5.3\pm1.3)\times10^5\;(25\%)$& $(7.8\pm3.9)\times10^6\;(51\%)$ \\
$f_{PAH}$& $0.9\pm0.2\;(25\%)$& $1.5\pm0.4\;(28\%)$& $1.8\pm0.4\;(25\%)$& $1.8\pm0.4\;(21\%)$ \\
$f_{PAH+}$& $0.8\pm0.1\;(15\%)$& $0.8\pm0.1\;(16\%)$& $0.6\pm0.1\;(21\%)$& $0.6\pm0.1\;(20\%)$ \\
$\alpha$& $2.2\pm0.5\;(24\%)$& $2.4\pm0.2\;(8\%)$& $2.4\pm0.1\;(4\%)$& $1.9\pm0.2\;(13\%)$ \\
$U_{Min}$& $5\pm2\;(36\%)$& $3\pm1\;(32\%)$& $2\pm1\;(29\%)$& $0\pm1\;(171\%)$ \\
$U_{Max}$& $654_{-654}^{+1863}\;(285\%)$& $36860\pm30208\;(82\%)$& $21481\pm23816\;(111\%)$& $601\pm521\;(87\%)$ \\
$A_V$& $5.7\pm0.2\;(3\%)$& $4.0\pm0.1\;(3\%)$& $3.9\pm0.1\;(3\%)$& $4.3\pm0.2\;(4\%)$ \\
$M_{oldstar}\;(M_\odot)$& $(1.1\pm0.0)\times10^{10}\;(3\%)$& $(3.8\pm0.1)\times10^9\;(3\%)$& $(3.8\pm0.1)\times10^9\;(3\%)$& $(2.8\pm0.1)\times10^9\;(4\%)$ \\
$f_{cold}$ (percentage) & $51\pm25\;(48\%)$& $80\pm11\;(14\%)$& $86\pm6\;(7\%)$& $99\pm1\;(1\%)$ \\
\hline
parameter & inner halo (5) & outer halo (6) & & \\
\hline
$M_d\;(M_\odot)$& $(2.5\pm0.9)\times10^5\;(35\%)$& $(1.5\pm0.7)\times10^5\;(45\%)$ & & \\
$f_{PAH}$& $1.5\pm1.7\;(117\%)$& $1.2\pm1.5\;(120\%)$ & & \\
$f_{PAH+}$& $0.4\pm0.3\;(76\%)$& $0.9\pm0.2\;(18\%)$ & & \\
$\alpha$& $2.5\pm0.1\;(2\%)$& $2.3\pm0.2\;(8\%)$ & & \\
$U_{Min}$& $2\pm2\;(110\%)$& $2\pm3\;(106\%)$ & & \\
$U_{Max}$& $18914\pm32890\;(174\%)$& $54711\pm47703\;(87\%)$ & & \\
$A_V$& $0.0\pm0.0\;(422\%)$& $0.0\pm0.1\;(148\%)$ & & \\
$M_{oldstar}\;(M_\odot)$& $(9.6\pm0.2)\times10^8\;(2\%)$& $(4.2\pm0.1)\times10^8\;(2\%)$ & & \\
$f_{cold}$ (percentage) & $93_{-16}^{+7}\;(17\%)$ & $84_{-22}^{+16}\;(26\%)$ & & \\
\hline
\end{tabular}
}
\end{minipage}
\end{table}
\end{landscape}

\section{Discussion}
\label{sec:discussion}

\subsection{The Dusty Disk of NGC 891}
\label{sec:dis_disk}

The changing morphology of the disk emission of NGC~891
as a function of wavelength, as
 indicated in Sect.~\ref{sec:iso_images}, nicely illustrates
the changing admixture of IR-emitting components --- from dominant
starlight at NIR wavelengths to PAHs and dust as the wavelength increases.
The PAH/dust distribution is complex, as shown in the maps
 (Figures \ref{fig:pah_cont} -
\ref{fig:other_cont}, \ref{fig:Apah} - \ref{fig:Awide}) as well as the  high resolution, high S/N
$\lambda\,$8 $\mu$m PAH  major axis profile
(Figure \ref{fig:77major_axis}).  The latter figure
 reveals new details of the PAH distribution
 along the major axis.

It is useful to compare the major axis profile of Figure
\ref{fig:77major_axis} to the molecular gas distribution. Scoville
et al. (1993) showed that the molecular gas in NGC~891 has a peak at
the center surrounded by a molecular ring located 3 - 7 kpc (65 -
150 arcsec) in radius, a structure also detected by Sofue \& Nakai
(1993). Our PAH profile, which reveals much more detail, follows
this general molecular gas distribution quite well, showing a
central peak and two peaks on either side at similar radii. In
addition, we detect an $\lambda\,$8 $\mu$m peak on the south-west
side of the disk at a radius of $\approx\,$ 140 arcsec (6.5 kpc)
which was seen in CO by Sofue \& Nakai (1993).  Thus, the dominant
correlation of the PAH emission in NGC~891 is with the molecular
gas, a result that is in agreement with Le Coupanec et al. (1999)
and also seen in other galaxies (e.g. Irwin \& Madden 2006; Regan et
al. 2006).  The much lower PAH emission at larger radii
 is likely due to lower quantities in the lower density extra-planar gas as well as
less excitation farther from SF regions.

Although our data were not of sufficient quality to obtain SEDs at
every point in the disk, we have obtained good results for four
locations as listed in Table \ref{tab:model_params}.


The ratio, $f_{PAH}$, is given in terms of the MW value of 0.046.
Thus, we can immediately compare the regions in NGC 891 to the MW in
this respect. For the four in-plane regions of NGC 891 this ratio
ranges from about 0.9 $\to$ 2, indicating that the PAH abundance is
roughly Galactic. Although the `centre' of this edge-on galaxy also
includes components along the line of sight, SED results for the
centre should nevertheless be weighted towards the true centre of
the galaxy and the fact that the central SED differs from the other
SEDs supports this view.  For example, we find that $f_{PAH}$ at the
centre of the galaxy is the lowest of the four in-plane regions.
This agrees with the results of Le Coupanec et al (1999) who found a
reduction in the $\lambda\,$11.3 $\mu$m PAH feature at the galactic
centre. It is also consistent with Smith et al. (2007) and Bendo et
al. (2008) who both find a decrease in PAH emission in the centres
of some galaxies. The fraction of dust exposed to low ISRFs ($f_{\rm
cold}$) is $\simeq50\%$ in the centre, where a significant fraction
of the dust emission comes from hotter dust, likely embedded in star
forming regions 
$f_{\rm cold}$ is between $\simeq80\%$ and $100\%$ in the disk, where the emission
seems to come primarily from a diffuse component.



Although the strength of the ISRF at lower wavelengths, $U_{min}$ is
approximately the same for all in-disk regions, surprisingly, the
heating intensity, $U_{max}$, is greatest in the south disk,
suggesting a greater range of heating environments than in the other
in-disk locations. As indicated in Sect.~\ref{sec:sed_results},
there is an enhancement in H$\,\alpha$ emission at this location and
the higher $U_{max}$ likely reflects the fact that the south
disk samples a region of star formation. These results also suggest
that the enhanced H$\,\alpha$ emission is likely real and not only
the result of dust obscuration elsewhere (see discussion in Kamphuis
et al. 2007). Note however, that the $U_{min}$ and $U_{max}$ values
are constrained only by the four data points in the FIR, and therefore
may reflect some of the noise in those values.

\subsection{Extraplanar Dust and PAHs in NGC 891}
\label{sec:dis_halo}

Previous investigations have shown that NGC 891 has extraplanar emission
 in neutral gas, DIG, molecular gas, and dust (eg. Oosterloo et al. 2007,
Fraternali et al. 2005, Rossa et al. 2004, Garcia-Burillo et al. 1992,
Howk $\&$ Savage 2000, see Sect.~\ref{sec:intro} 
and Table \ref{tab:halo_components}). Our results now show
that PAHs can be added to this list, with a measured $z$ height
extent of at least 2.3 kpc (Sect.~\ref{sec:minor_axis}). The $z$
width of the cold dust distribution at $\lambda\,$450 $\mu$m is
greater than those of both the warm dust ($\lambda\,$24 $\mu$m) as
well as PAHs
($\lambda\,$7.7 and 8 $\mu$m), as shown in
Figure~\ref{fig:77onk450_24} and noting error bars (see also
Figure~\ref{fig:77onHI850}).

Warm dust is only seen close to heating sources and
therefore it is not surprising that this
component has a narrower $z$ width than cold dust. 
  As for PAHs, we expect that PAHs may very
well exist at even higher $z$ latitudes than observed because we know
that PAHs are
present in H~I (see the SED models of Dwek et al. 1997, Draine \& Li
2007a, and Zubko et al. 2004 which show prominent PAH features associated
with the diffuse ISM) and the H~I halo of NGC~891 is extensive
(Table~\ref{tab:halo_components}).  The fact that
the PAH vertical width
 is closer to that of the warm dust (Figure~\ref{fig:77onk450_24})
  therefore suggests that 
the dominant
 PAH excitation sources are also embedded in the disk. 
Recent Galaxy Evolution Explorer observations 
have now revealed an FUV halo in NGC~891 (Gil de Paz et al. 2007)
which potentially provides a halo source of FUV photons. 
However, the FUV emission 
in galaxy halos is largely due
in-disk stellar continuum photons that are reflected from 
halo dust\footnote{There could also be a minor contribution from
shock excitation.}
(Hoopes et al. 2005).  In an optically thin halo environment, FUV photons
from the disk should dominate over scattered photons and, in either
case, the 
dominant excitation source ultimately resides from stars the disk.

 The extents of the extraplanar emission
 are summarized in Table \ref{tab:halo_components}. These
components are listed in order of largest distance from mid-plane to
smallest distance. All of the values given are for the farthest measured
extent of the halo (as opposed to a characteristic scale height) and
thus, they depend on whatever S/N is available for the data.  Therefore,
 Table \ref{tab:halo_components} is meant to provide a measure of the
full extent
of the extraplanar components in NGC 891 to the limits of current data.
Nevertheless,
Figures \ref{fig:77onk450_24} and \ref{fig:77onHI850}
display how the $z$ width of a subset of these quantities
compare to each other and the widths of these profiles agrees with the ordering
 of
Table \ref{tab:halo_components}.

\begin{table}
 \caption{Composition and Extent of Extraplanar Components in NGC 891.}
 \label{tab:halo_components}
\begin{tabular}{lll}
\hline
ISM component & distance from & reference \\
 & mid-plane (kpc) & \\
\hline neutral gas & 22 & Oosterloo et al. (2007) \\
 & 15 & Fraternali et al. (2005) \\
 & 5 & Swaters et al. (1997)\\
old stars & 3 & this paper (Figure \ref{fig:77onk450_24}) \\
DIG & 2.2 & Rossa et al. (2004) \\
 & 2 & Howk $\&$ Savage (2000) \\
warm dust & 2.5 & this paper (Figure \ref{fig:77onk450_24}) \\
cold dust & 2 & Alton et al. (1998) \\
PAHs & 2.3 & this paper (Figure \ref{fig:77onk450_24}) \\
molecular gas & 1 - 1.4 & Garcia-Burillo et al. (1992) \\
\hline
\end{tabular}
\end{table}

An important result of this work is our first attempt at a SED model in the
extraplanar (halo) region of an external galaxy
(Table~\ref{tab:model_params}).
The extra-planar dust mass results
are consistent with those determined from
extra-planar absorption (Howk $\&$ Savage 1997), after reasonable
corrections for different apertures are taken into account.
The $f_{PAH}$ ratios in the halo are similar to those in the disk. This result 
was unexpected given the observations of the vertical profiles in Figure 
\ref{fig:77onk450_24}, which show that the cool dust (at $\lambda\,$450 $\mu$m, 
for example) extends further into the halo vertically than the PAHs (at
$\lambda\,$7.7 $\mu$m, for example). However, the error bars are very large
for these halo results, and the SED model does not take into account the 
variation of the stellar populations from the disk to the halo.
A decrease of the hardness of the ISRF outward, would decrease the PAH strength, 
for a constant PAH-to-dust mass ratio, which is what we see here.
The quantity, $f_{cold}$, which is close to 1 in the halo,
is consistent with an interpretation in which few hot photons are available
for excitation.


While NGC 891 is known for large amounts of
neutral gas, DIG, and dust in its halo
(Sect.~\ref{sec:intro}), its PAH emission
 is not as extended in $z$ as has been seen in several
other galaxies which are {\it not} known for abundant halo gas and dust
(see Irwin $\&$ Madden 2006, Irwin et al. 2007).  Since the
excitation source for the PAHs appears to be within the disk
(see above), we speculate that FUV photons that could potentially
excite high latitude PAH emission in NGC~891 may instead be
absorbed by other gaseous and dust components.  That is,
other
components in the halo
of NGC~891 have some optical depth to FUV photons that originate
in the disk. 

\section{Conclusions}
\label{sec:conclusion}

We have examined 20 spatially resolved IR data sets for NGC~891, 14 of which are
newly reduced and/or newly presented data.  The emission spans wavelengths from
$\lambda\,$1.2 to 850 $\mu$m.  Although there are a variety of resolutions and S/N
for the different data sets, which include both ground-based and space-based observations,
we have compared the spatial distributions, both along and perpendicular to the major
axis for a subset of the data.  PAH emission is found to correlate well in the disk
with molecular gas, likely because of higher densities of gas and dust in these regions
as well as the abundant supply of UV photons from SF regions.

Of particular interest is the extraplanar, or `halo' gas and dust in
NGC~891.  The H~I, DIG, and dusty halos have been established by
others.  Our work has compared the $z$ extents of various components
and also shown that PAHs also exist in the halo of this galaxy to
heights of 2.3 $\pm$ 0.2 kpc at a 3$\sigma$ level.  NGC~891 is only
the 4th galaxy in which PAHs have been detected in the halo. We find
that cool halo gas, as measured by $\lambda\,$450 $\mu$m emission,
is broader in $z$ than both warm halo gas $\lambda\,$24 $\mu$m and
PAH emission, suggesting that the dust heating sources and PAH
excitation sources are within the disk.

We have used our multi-frequency data set to
construct complete IR spectra for four different locations in the disk of NGC~891
and two in its halo.  This has allowed us to model SEDs in different environments
both within
the disk of NGC~891 and, for the first time, in the halo.
The SED models have allowed us to
determine the mass of dust, fraction of PAHs, and fraction of
ionized PAHs for these different environments (Table~\ref{tab:model_params}).
We find that the PAH fraction, $f_{PAH}$, is similar to Galactic values (within a factor
of two) and this value is lowest in the galaxy's center, consistent with previous
studies of PAHs in galaxies.  The fraction of dust exposed to a colder
(i.e. equivalent to an equilibrium dust temperature less than 25 K) radiation field,
$f_{cold}$, is also lowest in the center, suggesting the presence of a hotter radiation
field at the center of the galaxy.

In the extraplanar dust, $f_{PAH}$ is similar to $f_{PAH}$ in the disk.
Therefore we cannot conclude a lower PAH abundance, especially since the SED
model used does not take into account a changing radiation field that is more typical
of a halo. The model used could take into account the changing radiation field, but
we would have needed additional constraints.   
The fraction of PAHs exposed to a cold radiation field is
very high in the halo and is almost unity.  Together with the vertical halo extents
indicated above, these results suggest that the halo PAHs are excited from escaping
disk photons.

\section*{Acknowledgements}
\label{sec:acknowledgements}

We thank Dr. R. Swaters for supplying us with the NGC 891 H I FITS
file, Dr. P. Alton for the SCUBA FITS files. This publication makes use 
of the ISO and Spitzer data archives from the European Space Agency and the
National Aeronautics and Space Administration (NASA), 
respectively. It also makes use of data products
from the Two Micron All Sky Survey, which is a joint project of the
University of Massachusetts and the Infrared Processing and Analysis
Center/California Institute of Technology, funded by NASA and the National Science
Foundation.

\newpage
\appendix
\section[]{}
\label{sec:appendix}

\begin{figure*}
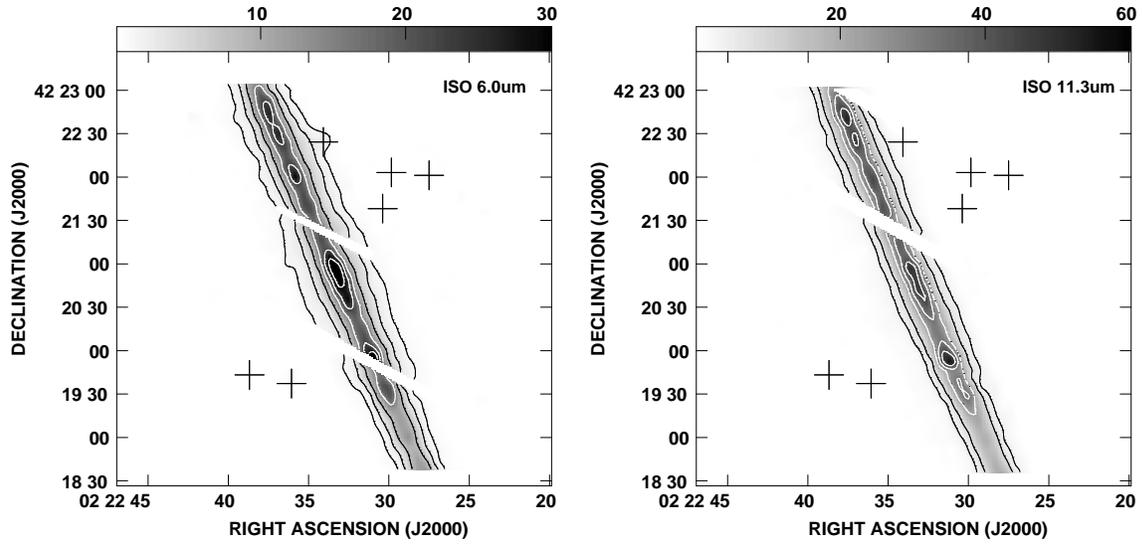

  \includegraphics[scale=0.4]{6_SELF_NEW.PS}
  \includegraphics[scale=0.4]{113_SELF_NEW.PS}
\caption{Surface brightness contour maps for remaining ISO bands that isolate PAH emission, where the others were shown in Figure \ref{fig:pah_cont}. 
Symbols as in Figure \ref{fig:pah_cont}. Contours are 0.3 (3$\sigma$), 0.8, 2, 4, 6, and 
8 MJy/sr for $\lambda\,$6.0, $\mu$m and 0.7 (3$\sigma$), 1.5, 4, 6, 8, 10 MJy/sr for $\lambda\,$11.3 $\mu$m.}
\label{fig:Apah}
\end{figure*}

\begin{figure*}
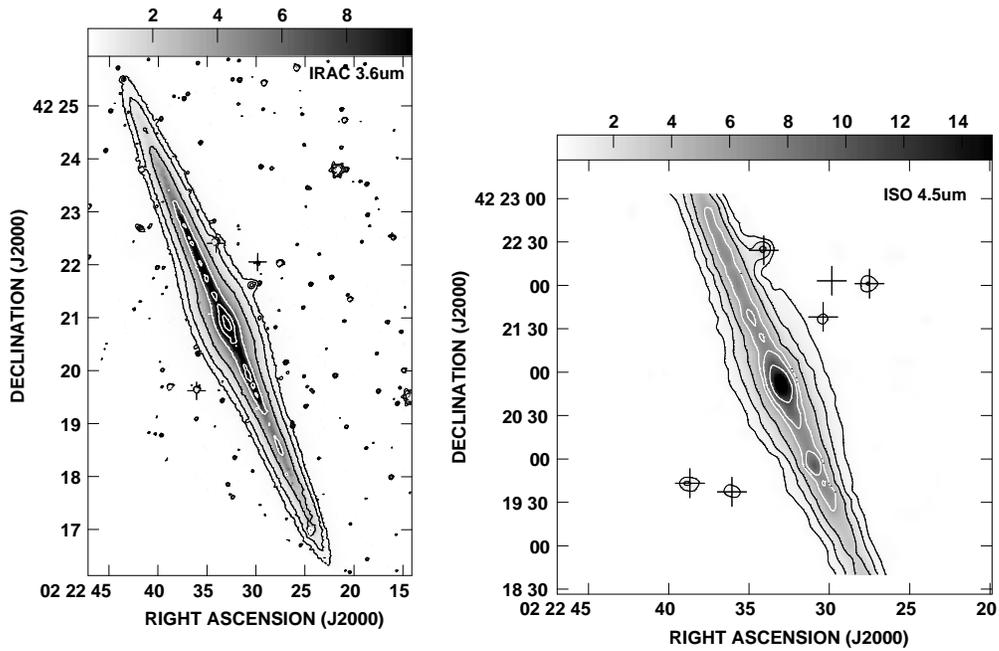
 
  \includegraphics[scale=0.4]{36_1_SELF.PS} 
  \includegraphics[scale=0.4]{45_SELF_NEW.PS}
\caption{Surface brightness contour maps for the remaining ISO and IRAC wave bands that isolates stellar emission, where the other was shown in Figure \ref{fig:stellar_cont}. 
Symbols as in Figure \ref{fig:pah_cont}. Contours are 0.08 (3$\sigma$), 0.15, 0.4, 1.2, 2.5, and 5 MJy/sr for
$\lambda\,$3.6 $\mu$m and 0.3 (3$\sigma$), 0.7, 1.5, 3, 4, 6 MJy/sr for
$\lambda\,$4.5 $\mu$m.}
\label{fig:Astar}
\end{figure*}

\begin{figure*}
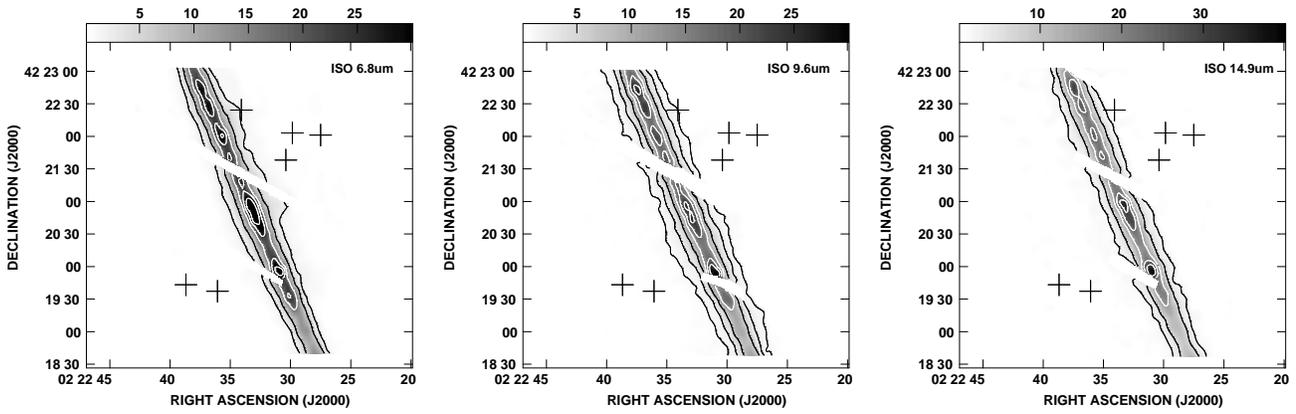

  \includegraphics[scale=0.3]{68_SELF_NEW.PS} 
  \includegraphics[scale=0.3]{96_SELF_NEW.PS} 
  \includegraphics[scale=0.3]{149_SELF_NEW.PS}
\caption{Surface brightness contour maps for the remaining ISO wave bands that isolate the MIR continuum, where the other was shown in Figure \ref{fig:other_cont}. 
Symbols as in Figure \ref{fig:pah_cont}. Contours are 0.5 (3$\sigma$), 
1.5, 4, 6, 8 MJy/sr for $\lambda\,$6.8 $\mu$m, 0.3 (3$\sigma$), 0.8, 2, 4, 6, 8 MJy/sr for $\lambda\,$9.6 $\mu$m, 
and 0.5 (3$\sigma$), 1.5, 4, 6, 8 MJy/sr for $\lambda\,$14.9 $\mu$m.}
\label{fig:Aother}
\end{figure*}
 
\begin{figure*}
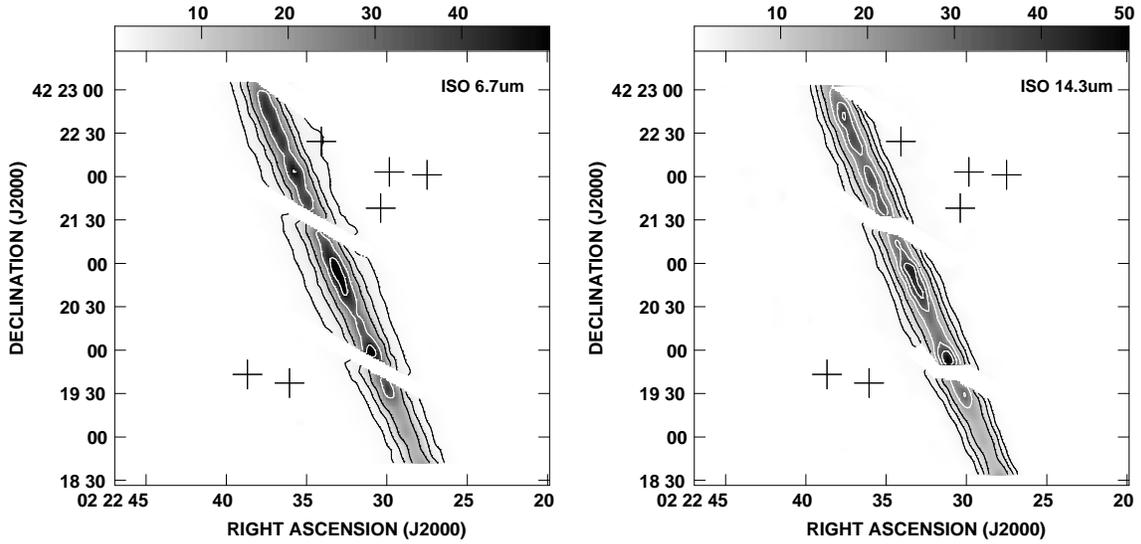

  \includegraphics[scale=0.4]{67_SELF_NEW.PS}
  \includegraphics[scale=0.4]{143_SELF_NEW.PS} 
\caption{Surface brightness contour maps for the ISO wide bands. 
Symbols as in Figure \ref{fig:pah_cont}. Contours are 0.3 (3$\sigma$), 0.8, 2, 5, 8 MJy/sr for 
$\lambda\,$6.7 $\mu$m and 0.5 (3$\sigma$), 1, 2, 4, 6, 8 MJy/sr for $\lambda\,$14.3 $\mu$m.}
\label{fig:Awide}
\end{figure*}

\begin{figure*}
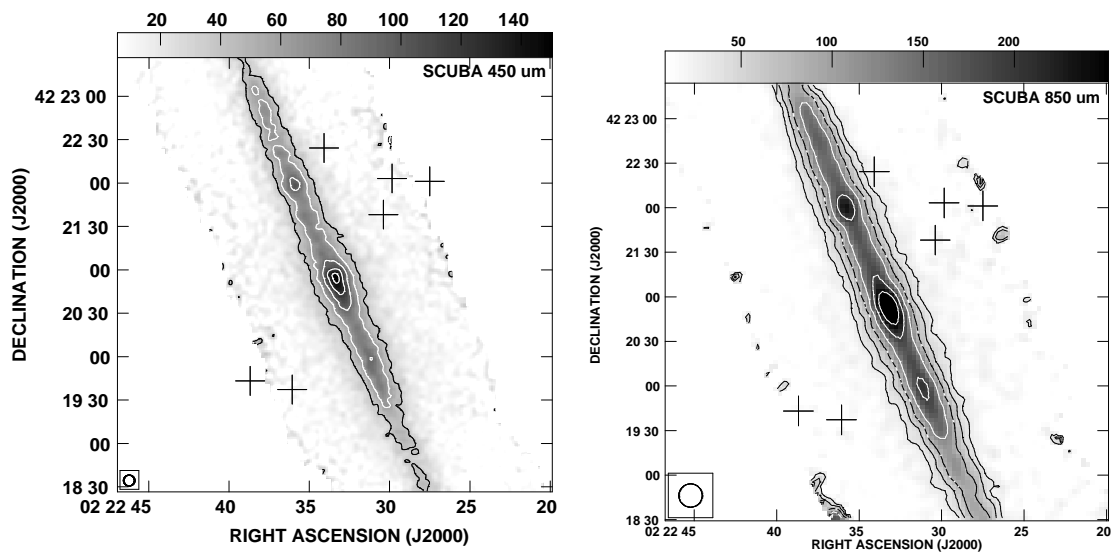

  \includegraphics[scale=0.4]{450_SELF.PS}
  \includegraphics[scale=0.38]{850_SELF.PS}
  \caption{The SCUBA
$\lambda\,$450 $\mu$m and $\lambda\,$850 $\mu$m
maps showing cool dust in NGC 891 (see Alton et al. 1998). Symbols
are as in Figure \ref{fig:pah_cont}.
For the $\lambda\,$450 $\mu$m map, the contours are
 2.5, 4, 6, 8, 9.5 Jy beam$^{-1}$.
For the $\lambda\,$850 $\mu$m map, the
contours are at  31 (3$\sigma$), 50, 80, 125, 190, and 250 mJy beam$^{-1}$.}
  \label{fig:cool_dust_map}
\end{figure*}

\bsp

\label{lastpage}

\end{document}